\documentclass[aps,pre,reprint,longbibliography,superscriptaddress,showpacs,amsmath,floatfix]{revtex4-1} 
\usepackage{mathtools}
\usepackage{amsmath,amssymb,mathrsfs,amsthm,physics}
\usepackage{mathrsfs}
\usepackage{bm}
\usepackage{bbm}
\usepackage{graphicx}
\usepackage{multirow}
\usepackage{color}
\usepackage{booktabs}
\usepackage{enumitem}
\usepackage{comment}
\usepackage{nameref}
\usepackage[colorlinks=true,linkcolor=blue,urlcolor=blue,citecolor=blue,anchorcolor=blue]{hyperref}
\usepackage{subcaption}  
\usepackage[font=small, justification=raggedright, singlelinecheck=off]{caption}
\usepackage{algorithm}
\usepackage{algpseudocode}
\usepackage{physics}

\definecolor{monbbleu}{RGB}{76, 114, 176}
\makeatletter
\renewcommand{\ALG@name}{Procedure}
\makeatother

\usepackage{tikz-cd}

\usepackage{tabstackengine}
\stackMath

\makeatletter
\g@addto@macro\normalsize{%
  \setlength\abovedisplayskip{5pt}
  \setlength\belowdisplayskip{5pt}
  \setlength\abovedisplayshortskip{5pt}
  \setlength\belowdisplayshortskip{5pt}
}
\makeatother

\usepackage{prettyref}
\newrefformat{SIsec}{SI~\ref{#1}}
\newrefformat{SIsubsec}{SI~\ref{#1}}
\setlist[itemize]{noitemsep, topsep=0pt}
\setlist[enumerate]{noitemsep, topsep=0pt}

\theoremstyle{definition}

\theoremstyle{remark}

\renewcommand{\vec}[1]{\boldsymbol{#1}}

\newcommand{\expect}[2][]{
    \mathbb{E}_{#1}\qty[#2]
}

\newcommand{\set}[1]{\left\{#1\right\}}
\newcommand{\mi}{I(G;X)}
\newcommand{\bigo}[1]{\mathcal{O}\left(#1\right)}

\DeclareMathOperator*{\argmax}{arg\,max}

\makeatletter
\newcommand{\labitem}[2]{%
\def\@itemlabel{\textcolor{black}{#1}}
\item
\def\@currentlabel{#1}\label{#2}}
\makeatother

\usepackage{mathtools}
\def\mutilsetcoeff#1#2{\ensuremath{\left(\kern-.3em\left(\genfrac{}{}{0pt}{}{#1}{#2}\right)\kern-.3em\right)}}

\begin{document}

\title{On the reconstruction limits of complex networks}

\author{Charles Murphy}%
\email[]{charles.murphy.1@ulaval.ca}
\affiliation{D\'epartement de physique, de g\'enie physique et d'optique, Universit\'e Laval, Qu\'ebec (Qc), Canada}
\affiliation{Centre interdisciplinaire en mod\'elisation math\'ematique de l'Universit\'e Laval, Qu\'ebec (Qc), Canada} 

\author{Ariane Lizotte}
\affiliation{D\'epartement de physique, de g\'enie physique et d'optique, Universit\'e Laval, Qu\'ebec (Qc), Canada}
\affiliation{Centre interdisciplinaire en mod\'elisation math\'ematique de l'Universit\'e Laval, Qu\'ebec (Qc), Canada} 

\author{François Thibault}
\affiliation{D\'epartement de physique, de g\'enie physique et d'optique, Universit\'e Laval, Qu\'ebec (Qc), Canada}
\affiliation{Centre interdisciplinaire en mod\'elisation math\'ematique de l'Universit\'e Laval, Qu\'ebec (Qc), Canada} 

\author{Vincent Thibeault}
\affiliation{D\'epartement de physique, de g\'enie physique et d'optique, Universit\'e Laval, Qu\'ebec (Qc), Canada}
\affiliation{Centre interdisciplinaire en mod\'elisation math\'ematique de l'Universit\'e Laval, Qu\'ebec (Qc), Canada} 

\author{Patrick Desrosiers}
\affiliation{D\'epartement de physique, de g\'enie physique et d'optique, Universit\'e Laval, Qu\'ebec (Qc), Canada} 
\affiliation{Centre interdisciplinaire en mod\'elisation math\'ematique de l'Universit\'e Laval, Qu\'ebec (Qc), Canada}
\affiliation{Centre de recherche CERVO, Qu\'ebec (Qc), Canada}

\author{Antoine Allard}
\email[]{antoine.allard@phy.ulaval.ca}
\affiliation{D\'epartement de physique, de g\'enie physique et d'optique, Universit\'e Laval, Qu\'ebec (Qc), Canada}
\affiliation{Centre interdisciplinaire en mod\'elisation math\'ematique de l'Universit\'e Laval, Qu\'ebec (Qc), Canada} 

\begin{abstract}
    Network reconstruction consists in retrieving the hidden interaction structure of a system from observations.
    Many reconstruction algorithms have been proposed, although less research has been devoted to describe their theoretical limitations.
    In this work, we take a first-principles approach and build on our earlier definition of \emph{reconstructability}—the fraction of structural information recoverable from data.
    We relate this quantity to the true data-generating (TDG) process and delineate an information-theoretic reconstruction limit, i.e., the upper bound of the mutual information between the true underlying graph and any graph reconstructed from observations. 
    These concepts lead us to a principled numerical method to assess the validity of empirically reconstructed networks, based on model selection and a quantity we introduce: the \emph{reconstruction index}. 
    This index approximates the reconstructability from data, quantifies the variability of the reconstructed network ensemble, and is shown to predict reconstruction error without requiring knowledge of the true underlying network.
    We characterize this method and test it on empirical time series and networks.
\end{abstract}

\maketitle

\section{Introduction}
\label{sec:introduction}

Complex systems, such as the brain, are naturally represented by complex networks that encapsulate intricate interactions between neurons or brain regions~\cite{bassett2017network, bassett2018nature, Lynn2019, sporns2020structure}.
Network representation unlocks a variety of tools with the potential to unravel not only brain functions and diseases~\cite{Supekar2008, hlinka2012using, forrester2020role}, but also gene expressions~\cite{wang2006inferring}, epidemics~\cite{peixoto2019network, prasse2020network} and the propagation of financial distress~\cite{musmeci2013bootstrapping}.
The main challenge is that such network representations are seldom measurable experimentally.
For example, the collected data are often indirect observations of the interactions, taking the form of counts of interactions or times series.
Moreover, these data are noisy, thereby making the network reconstruction task even more intricate~\cite{newman2018network, peixoto2018reconstructing, young2020bayesian, peel2022statistical}.

The task of reconstructing networks has been revisited many times, using different assumptions and approaches. 
Typically, network reconstruction is performed on multivariate time series~\cite{brugere2018network}, a procedure related to causal inference~\cite{eichler2012causal}. 
In this approach, we assume that the dynamics of the node activities is driven by some hidden network structure that we want to uncover.
Many heuristics have been proposed to perform network reconstruction from time series---involving scores like correlation~\cite{kramer2009network}, Granger causality~\cite{seth2005causal} or transfer entropy~\cite{schreiber2000measuring} between nodes---which are then thresholded to obtain a reconstructed network.
Other approaches proposed statistical frameworks to infer network from time series using graphical models~\cite{abbeel2006learning, montanari2009graphical, bresler2013reconstruction, amin2018quantum}, fully Bayesian models~\cite{peixoto2019network} and deep learning models~\cite{kipf2018neural}.

Another promising avenue for network reconstruction involves using pairwise observations for quantifying the uncertainty of empirical graphs.
%
In this setting, noisy pairwise observations are used to predict missing edges~\cite{clauset2008hierarchical, guimera2009missing, lu2011link}, estimate the edge uncertainty~\cite{newman2018network} and reconstruct the network altogether~\cite{peel2022statistical}. 
As for network reconstruction from time series, heuristics have also been considered for pairwise data (for example, in Ref.~\cite{goldberg2003assessing}). Recently, there has also been a resurgence in the interest towards Bayesian frameworks.
For instance, Ref.~\cite{young2020bayesian} proposed a general and Bayesian procedure to infer networks leveraging the conditional independence of the edges, which was then applied to a plant-pollinator network~\cite{young2021reconstruction}.
Reference~\cite{lizotte2022hypergraph} extended this framework to the reconstruction of hypergraphs with noisy observations and showed the benefit of including higher-order interactions for modeling pairwise measurements.
Other works used the modular structure of complex networks to improve the performance of their models~\cite{clauset2008hierarchical, guimera2009missing, peixoto2018reconstructing}.
To this date, the field of reconstruction of noisy networks remains a flourishing one.

As more technical progress is being made, more work is being dedicated to the theoretical challenges of network reconstruction.
For instance, Ref.~\cite{peel2022statistical} proposed a unifying framework for linking network data to network science theories, in which Bayesian network reconstruction is core and where they argue the suitability of the models is essential for network reconstruction.
Additionally, Ref.~\cite{prasse2022predicting} found that network reconstruction, on the basis of predicting the outcome of a deterministic
dynamical process, can lead to a wide range of networks. 
This aligns with the observations of Ref.~\cite{angulo2017fundamental} and earlier computational neuroscience findings~\cite{prinz2004similar} of network degeneracy~\cite{cropper2016consequences}, where diverse synaptic connection patterns can yield similar neuronal activity, illustrating the non-unique relationship between network structure and function.
%

An information-theoretic description of random networked processes has recently led to a broader understanding of this so-called structure-function relationship, linking predictability to reconstructability in complex networks through mutual information~\cite{murphy2024duality}.
This description revealed a duality between reconstructability and predictability, showing that in certain parameter ranges, an increase in predictability corresponds to a decrease in reconstructability, and vice versa.
In this work, we aim to further this theory on the network reconstruction front, providing an information-theoretic bedrock to such applications.

\begin{figure}
    \centering
    \includegraphics[width=.5\textwidth]{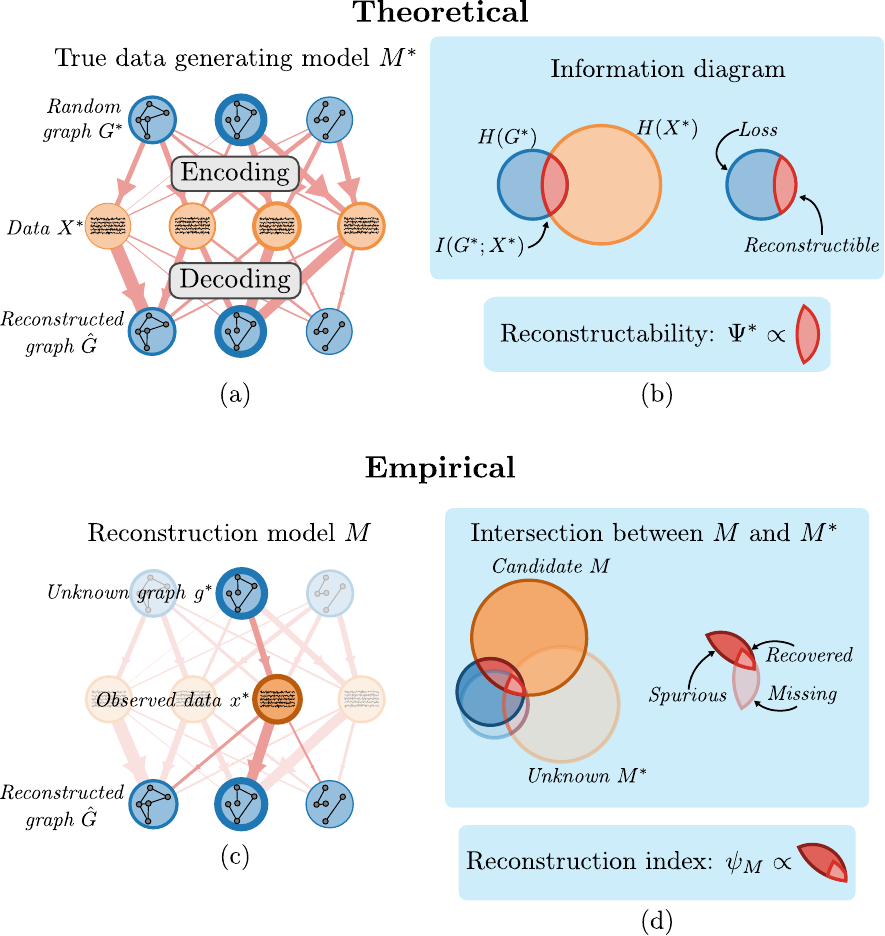}
    \caption{
    Illustration of the network reconstruction context from (a, b) a theoretical perspective and (c, d) an empirical perspective. 
    Panel (a) sketches how the true data generating model (TDG) $M^*$ operates, first by generating a graph, then by encoding it into the observations, and finally using these to decode---or reconstruct---the graph. 
    The thickness of the contour line around each graph and data example indicates the probabilities $P(G^*)$ (top and bottom layers) and $P(X^*)$ (middle layer). 
    The thickness of the edges connecting the graphs to the data illustrate the likelihood of the TDG $P(X^*|G^*)$, and those connecting the data to a reconstructed graphs, some distribution $P(\hat G|X^*)$.
    In panel (b), we illustrate in red the reconstructible information, utilizing an information-theoretic perspective.
    This information is part of the total information of $G^*$ and $X^*$---in blue and orange, respectively---and is also a fraction of the partial information of $G^*$ needed to completely reconstruct it (blue and red).
    Panels (c, d) show the analog of (a, b) when the model $M^*$ is unknown, where in panel (c) a single datum is accessible and reconstruction is done by a candidate model $M$, a priori different from $M^*$.
    In panel (d), we illustrate how $M$ and $M^*$ may overlap in the information they reconstruct---the information intersection (i.e., the correctly recovered information) and difference (i.e., the missing or spurious information).
    The reconstructability $\Psi^*$ and the reconstruction index $\psi_M$ are defined in subsection~\ref{subsec:recon} and subsection~\ref{subsec:recon_index}, respectively.
    }
    \label{fig:illustration}
\end{figure}

In the network reconstruction problem, our task is to infer a graph likely to have generated some observed data.
We first formally present this problem and the related mathematical concepts in Sec.~\ref{sec:network-reconstruction}.
Then, we revisit and adapt the framework of Ref.~\cite{murphy2024duality} in Sec.~\ref{sec:information-theoretic-limits}, allowing us to interpret the reconstruction problem in information-theoretic terms.
In doing so, we demonstrate the existence of an algorithm‑independent limit to network reconstruction---the reconstructability $\Psi^*$ [see Fig.~\ref{fig:illustration}(a,b)]---which bounds from above the mutual information between the true underlying network and the reconstructed one.
Inspired by this limit, we present and characterize in Sec.~\ref{sec:data-driven-reconstructability} a principled numerical method to assess the validity of reconstructed networks in an empirical setting (i.e., hidden generative process, one or few observations).
Our method is based on the reconstruction index, denoted $\psi_M$ for some reconstruction model $M$, which is an approximation of the reconstructability $\Psi^*$ that measures the dispersion of the reconstructed graph ensemble. 
The reconstruction index is shown to predict the reconstruction error \emph{without} knowing the true underlying graph [see Fig.~\ref{fig:illustration}(c,d)], assuming our modeling assumptions are aligned with the true underlying process. 
Finally, we apply our method to real systems in Sec.~\ref{sec:reconstructability-real-systems}.

\section{Network reconstruction}
\label{sec:network-reconstruction}
We formulate the network reconstruction problem following the illustration in Fig.~\ref{fig:illustration}(c). 
Let $g^*\in \mathcal{G}$ be some graph of $N$ nodes that represents the structure of the interactions between each pair of components in a system, where $\mathcal G$ is the set of all graphs of $N$ nodes.
The graph may be directed and weighted~\cite{peixoto2024network}, but we restrict our discussion to undirected and unweighted, for simplicity.
This graph structure is \textit{a priori} unknown to us, although it is indirectly observed through some data, denoted $x^*$, which may take any value in the set $\mathcal X$.
This data can take many forms---time series, pairwise measurements, etc.---and we assume it to be generated using $g^*$.
In what follows, we will further assume that $x^*$ is in fact a $N\times T$ matrix corresponding to $N$ coupled time series of length $T$, but we stress that our analysis may apply to any type of networked data.
The goal of network reconstruction is to infer the graph $g^*$ from the data $x^*$.

Taking a Bayesian perspective, the plausibility of a given graph $g\in\mathcal{G}$, given the observations $x^*$, is described by the posterior probability $P(G=g|X=x^*)$, i.e., the output of the Bayesian inference procedure.
A Bayesian reconstruction model is a generative process that consists of two discrete random variables $G$ and $X$, representing the graphs and the data respectively, and thus defines their joint probability mass function $P(G,X) = P(G) P(X|G)$, where $P(G)$ is the graph prior and $P(X|G)$, the data likelihood.
By virtue of Bayes' theorem, the posterior $P(G|X)$ is factored as follows:
\begin{equation}\label{eq:bayes}
  P(G|X) = \frac{P(X|G)P(G)}{P(X)}\,,
\end{equation}
where $P(X)$ is the normalization factor, called the evidence.

\subsection{Data generation process}
\label{sub:tdg-process}
\begin{figure}
  \centering
  \includegraphics[scale=0.9]{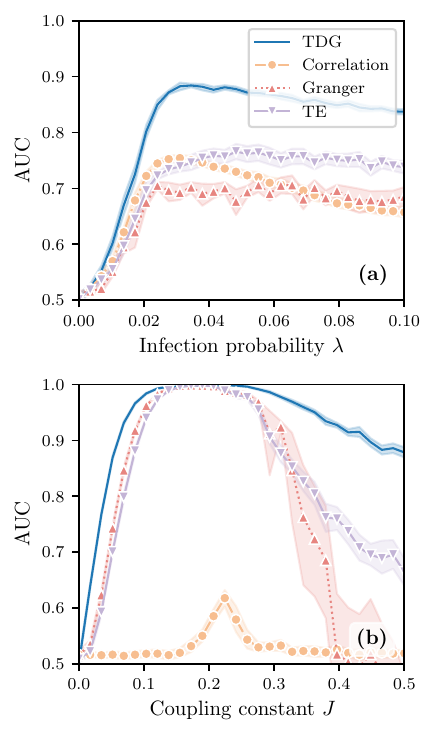}
  \caption{
  Performance comparison between the TDG model and heuristic reconstruction algorithms. 
  In both panels, we show the area under the receiver operating characteristic curve (AUC) of the reconstruction models as a function of a parameter of the model that generated the data: (a) the Susceptible-Infection-Susceptible (SIS) dynamics and (b) Glauber dynamics (see Table~\ref{tab:dynamics} for the definitions of the dynamics). 
  We generated graphs of $N=100$ nodes with the Erd\H{o}s-R\'enyi model (Eq.~\eqref{eq:erdos-renyi}), where the number of edges is $E=250$. 
  We also generated time series of $T=500$ time steps; the parameters other than the infection probability $\lambda$ and the coupling constant $J$ (which are fixed within the likelihood during the inference of the TDG) are specified in Table~\ref{tab:dynamics}.
  Each data point corresponds to the AUC average over 24 reconstruction experiments, each experiment with different realizations of $G^*$ and $X^*$, and the shaded regions around the points show a 90\% confident interval from the mean.
  For further technical details, see Sec.~\ref{sub:tdg-process}.
  }
  \label{fig:tdg-vs-heuristics} 
\end{figure}

A Bayesian reconstruction model, composed of the two random variables $G$ and $X$, reflects our assumptions about how the unobserved graph and observed data came to be.
In other words, the model $M = (G, X)$ represents a generative process for the pairs $(g^*, x^*)$ [see Fig.~\ref{fig:illustration}(a)].
Accordingly, there are many reconstruction models that may describe the data to various degrees of correctness.
Throughout this work, we assume the existence of a unique generative process, referred to as the true data-generating (TDG) model $M^* = (G^*, X^*)$, which \emph{truly} produced the graph $g^*$ and the observed data $x^*$ with probabilities $P(G^* = g^*)$ and $P(X^* = x^* | G^* = g^*)$, respectively.
In turn, any reconstruction model may be described by a reconstructed random graph $\hat G$, that depends on $X^*$.
The complete process consisting of the graph and data generation followed by the reconstruction of the graph is therefore described by the random variable triplet $(G^*, X^*, \hat G)$, whose joint probability distribution is
\begin{align}\label{eq:generation_reconstruction}
    P(G^*, X^*,\hat G) = P(G^*)P(X^*|G^*)P(\hat G|X^*)\,.
\end{align}
In general, the distribution $P(\hat G|X^*)$ may be any distribution over $\mathcal{G}$, but for Bayesian models such as $M$, it is precisely given by the posterior of $M$:
\begin{equation}\label{eq:hatg_vs_g}
    P(\hat{G}=g|X^*=x^*) = P(G=g|X=x^*)\,
\end{equation}
for all $g\in\mathcal{G}$, such that $P(G|X)$ is given by Eq.~\eqref{eq:bayes}.
Note that the reconstructed random graph $\hat G$ and the random graph $G$ of model $M$ conceptually describe two different quantities, although they are related through Eq.~\eqref{eq:hatg_vs_g}.
Indeed, $\hat G$ appears in the reconstruction process involving the TDG and $G$ is part of a completely separate generative process.
In other words, $\hat G$ depends explicitly on $M^*$, via $P(\hat G| X^*)$, whereas $M$ is independent from it (i.e., $P(G, X|G^*, X^*) = P(G, X)$).
The consideration that $\hat G$ is, in fact, resulting from a Bayesian procedure through a generative model $M$, instead of any---potentially nongenerative---algorithm such as the inverse correlation method \cite{kramer2009network}, will prove useful in the following sections.

From an information-theoretic perspective, data generation encodes information about the graph $G^*$ into potentially noisy observations $X^*$, while network reconstruction decodes these observations back into a graph $\hat G$ as shown in Fig.~\ref{fig:illustration}(a, b).
The encoding of $G^*$ into $X^*$ is generally lossy, meaning that only a fraction of its information can be recovered; the rest being lost in the process.
In turn, any reconstruction model $M$ different in distribution from $M^*$ may therefore recover a fraction of the reconstructible information while potentially introducing spurious information through their inductive biases [see Fig.~\ref{fig:illustration}(c,d)], resulting in a degradation of performance.

This is well shown in Fig.~\ref{fig:tdg-vs-heuristics} through reconstruction performance, where we used two synthetic TDG processes to compare the TDG reconstruction model performance with that of three heuristic reconstruction algorithms.
In this experiment, we sample the true graph with probability
\begin{equation}\label{eq:erdos-renyi}
  P(G^*) = \binom{\binom{N}{2}}{E}^{-1}\,,
\end{equation}
which corresponds to the Erd\H{o}s-R\'enyi (ER) model, with $E$ being the (given) number of edges in the graph; and sample time series from the Susceptible-Infected-Susceptible (SIS) model in panel (a) and Glauber models in panel (b). 
For further details regarding the graph and data models, see Appendices~\ref{app:graph-models} and \ref{app:markov-chain}.
Then, assuming a given reconstruction model, we compare the reconstructed graph with original one using the area under the receiver operating characteristic curve (AUC) to measure reconstruction performance.
We repeat this set up for many parameter values to populate the AUC performance curves in Fig.~\ref{fig:tdg-vs-heuristics}.
As a comparison, we use three different well-known reconstruction algorithms: the correlation matrix method~\cite{kramer2009network},  Granger causality method~\cite{schreiber2000measuring} and the transfer entropy method~\cite{seth2005causal} (see Appendix~\ref{app:heuristics} for details).
The results in Fig.~\ref{fig:tdg-vs-heuristics} show quite unambiguously and unsurprisingly that the TDG model outperforms the reconstruction heuristics.

Yet, even the TDG reconstruction model cannot reconstruct the graph perfectly.
For instance, in Fig~\ref{fig:tdg-vs-heuristics}(b), the AUC of the TDG model tends to $\frac12$---equivalent to random guessing---when the coupling also goes to zero.
In this scenario, $X^*$ and $G^*$ are independent and it is actually impossible to reconstruct the graph, since any graph could have generated the data with the exact same probability.
The same phenomenon occurs to a lesser extent for the other coupling values as well as for the SIS dynamics, where the TDG model performance is imperfect for every infection probability.
These imperfections are attributed to the lost information in the encoding of $G^*$; no model can extract more information than what is contained in $X$.
In practice, the encoding's loss stems from many sources, for example noise in the dynamics and degeneracy, where many networks lead to similar dynamics.
The degeneracy phenomenon is well-established in computational neuroscience~\cite{prinz2004similar,cropper2016consequences} and has more recently appeared in network science~\cite{prasse2022predicting} too.
A reconstruction limit independent of the reconstruction algorithm clearly exists, where a perfect reconstruction is simply not attainable even in the best-case scenario.
This is a key insight that we will explore in the following sections (especially Sec.~\ref{sec:data-driven-reconstructability}).

\subsection{Reconstructing a single edge}
\label{sec:reconstructing-single_edge}
To gain better intuition about this reconstruction limit, we consider the reconstruction of a graph that may only contain a single edge.
Let $G^*$ be a random graph of two nodes, that may be connected by a single edge with probability $p$, and disconnected with probability $1-p$.
This edge is observed through a noisy process $X^* = (X_1, ..., X_T)$ with $T$ time steps, where $X_i$ is a binary variable that takes the value $1$ if the edge has been observed and $0$ otherwise.
We assume that the noisy process can induce true positives and false positives, each with known probabilities $q$ and $r$, respectively, making the reconstruction problem more challenging.
The likelihood $P(n|a)$ that the edge has been observed $n$ times, given that it is present ($a=1$) or not ($a=0$), is a binomial distribution:
\begin{equation}
  \begin{split}
    P(n|a) = \binom{T}{n}&\Big[aq + (1 - a)r\Big]^n\\&\Big[1-aq - (1-a)r\Big]^{T-n}\,.
  \end{split}
\end{equation}
Note that this model possesses a symmetry where interchanging $q$ and $r$ and mapping $a\to 1-a$ leaves the likelihood invariant.
However, we avoid this non-identifiability issue by not inferring $p$ and $r$.
 
To calculate the posterior probability of the edge being present, we find the evidence of the data
\begin{align}
  P(n)
    & = \sum_{a=0}^1 P(n|a) P(a) \nonumber \\
    & = \binom{T}{n} q^n (1-q)^{T-n} \Big[p + \eta^{T-n}\lambda^n(1 - p)\Big]\,,
\end{align}
where
\begin{equation}
  \lambda = \frac{r}{q}\quad\text{and}\quad \eta = \frac{1 - r}{1 - q}\,.
\end{equation}
This leads to the posterior probability of the edge being present
\begin{equation}
  P(a=1|n) = \frac{p}{p + \eta^{T-n}\lambda^n(1 - p)}\,.
\end{equation}

\begin{figure}
  \centering
  \includegraphics[scale=0.9]{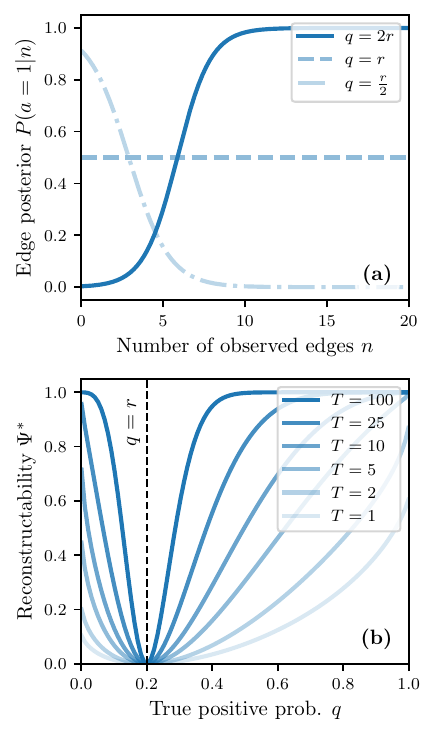}
  \caption{
  Posterior probability of a reconstructed edge: (a) Posterior versus the number of times $n$ the edge has been observed, (b) reconstructability of the edge versus $q$. 
  In panel (a), we fixed the number of observations $T = 20$, the prior edge occupancy probability $p = \frac12$ and the false positive probability $r = 0.2$. 
  We varied the true positive probability such as $q \in \set{2r, r, \frac{r}{2}}$  (solid, dashed and dotted lines, respectively).
  In panel (b), we show the reconstructability curves for different numbers of observations $T$ as indicated in the legend.
  The vertical dashed line indicates the value of $q$ for which the edge is not reconstructable, i.e., when the true positive and false positive probabilities are the same---i.e., $q = r$.
  }
  \label{fig:single-edge}
\end{figure}
Figure~\ref{fig:single-edge} shows the behavior of $P(a=1|n)$ when varying the number $n$ of times the edge is observed and the true positive probability $q$.
Assuming that $r < q$, observed edges are mostly true positives and thus the edge is predicted to exist if $n$ is sufficiently large; otherwise, it is not since the expected number of true and false positives don't match the observations.
Conversely, if $r>q$, then most observed edges are false positives, meaning that $g^*$ is more likely to contain an edge when $n$ is small.
Interestingly, the edge becomes more challenging to reconstruct as $q$ gets closer to $r$, where the probability to reconstruct the edge approaches $\frac12$ (see Fig.~\ref{fig:single-edge}(a)).
In this regime, $a$ and $1-a$ are interchangeable and it becomes impossible to tell if the edge exists or not---any attempt at reconstructing this graph would be unfruitful.
This is precisely the intuition we want to capture with the reconstruction limit: When is there enough information to properly reconstruct the structure, or to what extent is a system's structure reconstructible?
In the next section, we present an information-theoretic framework that quantifies this limit.

\section{Information-theoretic reconstruction limits}
\label{sec:information-theoretic-limits}
As discussed above, we can think of the TDG process $X^*$ as a noisy encoding of the true graph $G^*$.
This amount of encoded information is fundamentally limiting our ability to reconstruct $G^*$ accurately; it is impossible to recover more information than what is contained in the data.
This also means that the limit is independent of the reconstruction models or algorithms.
Any reconstruction algorithm therefore aims to extract as much of the encoded information as possible, some being more efficient than others.

\subsection{Entropy}
Our goal is to formalize this intuition in information-theoretic terms.
In information theory, information is related to the concept of entropy, which measures the uncertainty of a random variable.
For a random variable $G$, the entropy $H(G)$ is expressed as
\begin{align}
  H(G)
    & = -\expect[G]{\log P(G)} \nonumber \\
    & = -\sum_{g\in\mathcal{G}} P(G=g)\log P(G=g)\,.
\end{align}
The entropy $H(G)$ measured in bits (assuming $\log(x) \equiv \log_2(x)$, which will henceforth be the case) quantifies the minimal number of binary questions one needs to answer, on average, to perfectly identify the graph generated by $G$.
When $H(G) = 0$, the random variable $G$ can only yield one graph, meaning that $P(G=g) = 1$ for some $g$.
One can also measure the conditional entropy of a random variable $G$, given another random variable $X$, as
\begin{align}
  H(G|X)
    & = -\expect[X,G]{\log P(G|X)} \nonumber \\
    & = -\sum_{g\in\mathcal{G}} \sum_{x\in\mathcal{X}} P(G=g, X=x) \nonumber \\
    &  \qquad \qquad \qquad \times \log P(G=g | X = x)\,.
\end{align}
Like $H(G)$, $H(G|X)$ also measures uncertainty, but this time assuming that $X$ is known.
In Bayesian terms, $H(G)$ is the entropy of the prior $P(G)$, while $H(G|X)$ is the entropy of the posterior $P(G|X)$.

\subsection{Network reconstructability}
\label{subsec:recon}
Those information-theoretic tools can be used to define the reconstruction limit.
Consider the mutual information between the true and the reconstructed random graphs
\begin{equation}
    I(G^*; \hat G) = H(G^*) - H(G^*|\hat G)\,,
\end{equation}
where
\begin{equation}
    H(G^*|\hat G) = -\expect[G^*, \hat G]{\log P(G^*|\hat G)}\,
\end{equation}
is the entropy of the true graph given the reconstructed one. 
The conditional probability $P(G^*|\hat G) = P(G^*,\hat G)/P(\hat G)$  is such that both $P(G^*, \hat G)$ and $P(\hat G)$ are marginal distributions of $P(G^*, X^*, \hat G)$ [Eq.~\eqref{eq:generation_reconstruction}].
Three observations regarding this performance measure are in order.
First, the quantity $I(G^*;\hat G)$ may be interpreted as measuring the similarity between the information contents of $G^*$ and $\hat G$.
The higher it is, the more similar $G^*$ and $\hat G$ are and the better is the reconstruction.
Conversely, when $I(G^*;\hat G)=0$, it is minimized and both graphs are independent from one another.
Note that similar mutual information measures have been used as a performance measure in the context of community detection for comparing pairs of partitions \cite{newman2020improved,jerdee2024mutual}.

Second, $I(G^*;\hat G)$ is related to the probability of error, defined as
\begin{equation}
    p_e = P(\epsilon)\,,
\end{equation}
where $\epsilon = \mathbb{I}[G^* \neq \hat G]$, with $\mathbb{I}[\cdots]$ being the indicator function, denotes a binary random variable that takes the value 1 when $G^* \neq \hat{G}$ and 0 otherwise.  
This relationship can be shown through Fano's inequality \cite{cover2006elements}:
\begin{equation} \label{eq:binary-entropy-first-appearance}
    H(G^*|\hat G) \leq h(p_e) + H(G^*) p_e\,,
\end{equation}
where $h(p) \equiv -p\log p - (1 - p)\log(1 - p)$ is the binary entropy. 
Indeed, given that $h(p_e) \leq 1$, modifying Fano's inequality yields
\begin{equation}
    p_e \geq 1 - \frac{I(G^*;\hat G) + 1}{H(G^*)}\,.
\end{equation}
This lower bound on the probability of error is minimized when $I(G^*;\hat G)$ is maximized.

Third, using the data processing inequality~\cite{cover2006elements}, it is also related to the mutual information between $G^*$ and $X^*$ as follows:
\begin{equation}
    I(G^*; \hat G) \leq I(G^*;X^*)\,,
\end{equation}
where the mutual information upper bound is expressed as
\begin{align}\label{eq:mutual-information}
  I(G^*;X^*) = H(G^*) - H(G^*|X^*)\,
\end{align}
is the mutual information between the true graph $G^*$ and the data process $X^*$. 
Intuitively, $I(G^*;X^*)$ quantifies the amount of reconstructible information that both $X^*$ and $G^*$ share---i.e., the amount of information that $X^*$ contains about $G^*$ [see Fig.~\ref{fig:illustration}(b)].
The mutual information $I(G^*;X^*)$ also sets the maximum in reconstruction performance as measured by $I(G^*; \hat{G})$: It is the reconstruction limit.

The mutual information $I(G^*;X^*)$ is itself bounded between $0$ and $H(G^*)$~\cite{cover2006elements}.
When $I(G^*;X^*) = 0$, $X^*$ and $G^*$ are independent and thus the data $X^*$ contains no information about the graph $G^*$.
In turn, it is impossible for any reconstruction model $M$ to extract information from the data, regardless of its specification.
When $I(G^*;X^*) = H(G^*)$, the data $X^*$ contains all the information about the graph $G^*$.
In this case, it is in principle possible to perfectly reconstruct the graph without any error, assuming the model $M$ is optimal, i.e., it can extract all the available information. 
We will further explore this notion of optimality in Sec.~\ref{sec:reconstructability-performance}.

Since the value of $I(G^*;X^*)$ depends on the amount of information $H(G^*)$ that needs to be extracted, it is easier to reason about it in terms of proportions.
Thus, we define the \emph{reconstructability} $\Psi^*$ of $G^*$ from $X^*$ as the uncertainty coefficient
\begin{equation}\label{eq:reconstructability}
  \Psi^* = \frac{I(G^*;X^*)}{H(G^*)}\,.
\end{equation}
The reconstructability has been described thoroughly in Ref.~\cite{murphy2024duality} and helped unveiling a special duality between our ability to predict the time evolution of a system and our ability to reconstruct the interactions between its constituents.
As it is a normalized version of the mutual information upper bound $I(G^*;X^*)$, the reconstructability is bounded between $0$ and $1$.
When $\Psi^* = 0$, any attempt at reconstruction is futile, whereas it is theoretically possible to decode all the information when $\Psi^* = 1$.
As such, the reconstructability is a measure of the average proportion of information that can be extracted from the data about the graph.
For instance, when it is equal to $\frac12$,  it precisely means that half of the graph information is, on average, contained in the data and that in turn half of it can possibly be reconstructed.
We stress that $\Psi^* = \frac12$ may not be directly interpreted as half of the graph's edges being reconstructible.
Rather, information may generally be distributed in a heterogeneous way over the graph's structure, as a single bit of information may reconstruct more than one edge in the graph depending on how correlated they are.

Going back to the earlier example of a single edge of Sec.~\ref{sec:reconstructing-single_edge}, we can perform the complete calculation analytically. 
First, we can calculate the entropy of the prior,
\begin{equation}
  H(G^*) = h(p)\,,
\end{equation}
recalling that $h(p)$ is the binary entropy defined below Eq.~\eqref{eq:binary-entropy-first-appearance}, and the entropy of the posterior is
\begin{align} \label{eq:edge-posterior-entropy}  
  H(G^*|X^*)
    & = - \sum_{n=0}^T \sum_{a=0}^1 P(a|n) P(n) \log P(a|n) \nonumber \\
    & = \sum_{n=0}^T \binom{T}{n} q^n(1 - q)^{T-n}\left[p + \eta^{T-n}\lambda^n(1 - p)\right] \nonumber \\
    & \qquad \qquad \times h\left(\frac{p}{p + \eta^{T-n}\lambda^n(1 - p)}\right)\,.
\end{align}
This leads to the reconstructability of the edge using Eq.~\eqref{eq:reconstructability}, which is plotted in Fig.~\ref{fig:single-edge}(b).
As expected, the reconstructability typically increases as the number of observations $T$ increases, even reaching 1 in some cases (e.g., when $q\to1$).
Also, notice how the reconstructability is zero for every value of $T$ when the true positive probability $q$ is equal to false positive probability $r$.
This shows that as true positives and false positives become indistinguishable, the edge becomes impossible to reconstruct.

\subsection{Optimal reconstruction performance}
\label{sec:reconstructability-performance}

The reconstruction limit corresponds to the maximum performance, as measured by $I(G^*;\hat G)$, achievable by any algorithm.
Hence, any reconstruction model that is capable of reaching this limit, i.e., $I(G^*;\hat G) = I(G^*;X^*) = \Psi^* H(G^*)$, is said optimal in its reconstruction abilities.
It is not surprising that the TDG model is optimal according to this definition.
As a result, the reconstructability $\Psi^*$ can also be interpreted as a reconstruction performance measure of the TDG model $M^*$.

In fact, the reconstructability relates to standard performance measures.
One such example is the \emph{posterior loss}---also known as the log loss and the cross-entropy loss in the machine learning community. 
This measure is defined as 
\begin{align}
  \mathcal{L}\Big(\vec{a}^*, \vec{\pi}^*(x)\Big) = -\textstyle{\sum_{i<j}}&\left[a_{ij}^*\log \pi_{ij}(x)\right.\label{eq:ce-loss}\\ &\left.+ (1 - a_{ij}^*)\log (1 - \pi_{ij}(x))\right]\,,\nonumber
\end{align}
where $\vec{a}^*$ denotes the adjacency of the true graph $g^*$, such that $a_{ij}^*$ counts the number of edges connecting the nodes $i$ and $j$ (we use the convention that $a_{ii}^*$ is always a multiple of 2) and $\vec{\pi}(x) = [\pi_{ij}(x)]_{ij}$ is the predicted matrix of the posterior marginal probabilities of the edge occupancy for some model $M$.
We show in Appendix~\ref{app:posterior-loss} that, provided that the data is generated with $M^*$ and the reconstruction is performed with $M$, the posterior entropy and the expected posterior loss are equal if $M$ is equal to $M^*$ in distribution.
Consequently, the reconstructability is linearly related to the posterior loss as follows:
\begin{equation}\label{eq:linear-relationship-u-celoss}
    \Psi^* \approx 1 - \frac{\expect[G^*, X^*]{\mathcal{L}\Big(\vec{A}^*, \vec{\pi}(X^*)\Big)}}{H(G^*)}\,,
\end{equation}
where $\vec{A}^*$ is the random 
adjacency matrix of $G^*$.

\begin{figure}
  \centering
  \includegraphics[width=0.45\textwidth]{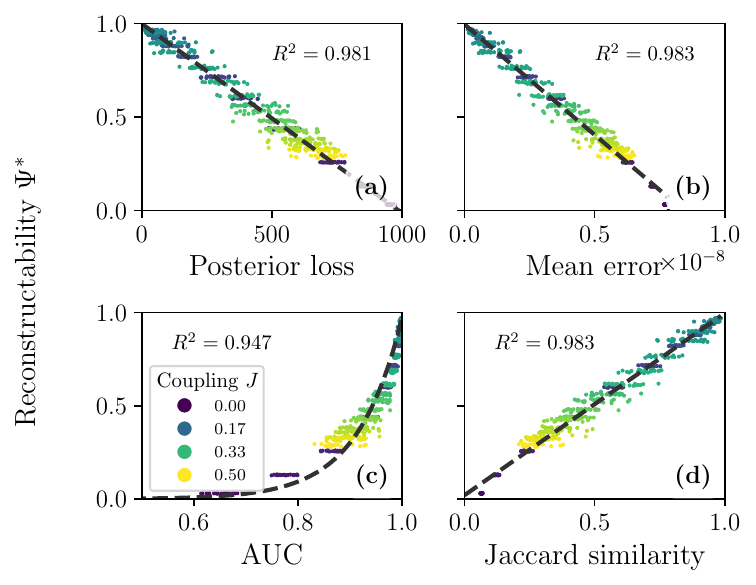}
  \caption{
      Comparison between reconstructability and different performance metrics: (a) posterior loss (Eq.~\eqref{eq:ce-loss}), (b) mean error $\binom{N}{2}^{-1}\sum_{i<j} |a_{ij} - \pi_{ij}(x)|$, (c) area under the receiver operating characteristic curve (AUC) and (d) Jaccard similarity (see Ref.~\cite[Eq.~11]{peixoto2019network}). 
      Each point shows a different realization of the Glauber dynamics whose graphs are generated from the Erd\H{o}s-Rényi model with $N=100$ nodes and $E=250$ edges, and whose initial conditions are random.
      Reconstructions are performed with the same model, whose parameters are fixed to those used for generating the data.
      We used time series of $T=500$ time steps (as in Fig.~\ref{fig:tdg-vs-heuristics}, the parameters other than the coupling constant $J$ are specified in Table~\ref{tab:dynamics}). 
      We generated 24 realizations of the process for each value of $J$ and used 30 different coupling values uniformly spaced between 0 and 0.5. 
      These coupling values are fixed during inference.
      The colors indicated in the legend show the value of $J$ associated with the point (only 6 colors are shown for conciseness).
      Finally, we show the determination coefficients $R^2$ relating the performance metrics to $\Psi^*$ in each plot. 
      For panel (a), we used Eq.~\eqref{eq:linear-relationship-u-celoss} directly to evaluate the determination coefficient, and for panels (b) and (d), we used standard linear regression to find the slope and estimate $R^2$. 
      For panel (c), because the scaling is not linear like the other cases, we used instead log-linear regression to estimate $R^2$.
  }
  \label{fig:recon-vs-metrics}
\end{figure}

Figure~\ref{fig:recon-vs-metrics} shows further numerical evidence of the relationship between the reconstructability of the Glauber model and reconstruction performance measures, including the posterior loss.
In Figs.~\ref{fig:recon-vs-metrics}(a) and (b), we show how $\Psi^*$ is well correlated with metrics quantifying the recontruction error, such as the posterior loss and the mean error.
Similarly, Figs.~\ref{fig:recon-vs-metrics}(c) and (d) show that the reconstructability is positively correlated with the area under the receiver operating characteristic curve (AUC) and the Jaccard similarity~\cite{peixoto2019network}, both measuring the similarity between the true and reconstructed graphs.

\subsection{Reconstructability of hierarchical Bayesian models}
\label{ssec:reconstructability--hierarchical}
Hierarchical models may be used for network reconstruction where additional parameters, namely the random variables $\theta$ and $\phi$, are included to parametrize the prior and likelihood respectively.
In this case, the likelihood $P(X|G, \phi)$ of the model $M$ depends on some unknown parameters $\phi$ with prior $P(\phi)$ and the graph prior $P(G|\theta)$ depends on other unknown hyperparameters $\theta$ with hyperprior $P(\theta)$.
During network reconstruction, hyperparameters $\theta$ are inferred jointly with $G$, as they are included in the posterior distribution of the model, while the parameters $\phi$ are marginalized as follows:
\begin{align}\label{eq:hyperparameter-posterior}
    P(G, \theta|X) &= \sum_{\varphi\in\Phi} \frac{P(X|G, \phi=\varphi) P(\phi=\varphi) P(G|\theta) P(\theta)}{P(X)}\,,
\end{align}
where $\phi$ and $\theta$ are assumed independent. 
Note that the sum becomes an integral over the corresponding probability density functions where $\phi$ is continuous, such that $\rho(\phi)$ is its prior density.
In this section, we show how our framework can used on such hierarchical models, without any modification.

Consider the case where a TDG model with variables $(G^*, X^*)$ also includes hyperparameters, denoted $\theta^*$ with probability distribution $P(\theta^*)$, such that $G^*$ is conditioned on $\theta^*$---i.e., $P(\theta^*, G^*) = P(\theta^*) P(G^*|\theta^*)$.
Let $\hat\theta$ and $\hat G$ be the reconstructed random parameters and graph, respectively, which are reconstructed from $X^*$ via some distribution $P(\hat\theta, \hat G|X^*)$.
The random variables $(\theta^*, G^*)$ are related to those of the reconstruction model $(\hat \theta, \hat G)$ via $X^*$ as follows:
\begin{equation}
    P(\theta^*, G^*, X^*, \hat \theta, \hat G) = P(\theta^*, G^*)P(X^*|G^*) P(\hat \theta, \hat G|X^*)\,,
\end{equation}
where, again assuming that we use a Bayesian reconstruction model $M$, we let $P(\hat \theta=\vartheta, \hat G=g|X^*=x^*) = P(\theta=\vartheta, G=g|X=x^*)$, which is given by Eq.~\eqref{eq:hyperparameter-posterior}.
In this case, the mutual information between $(\theta^*, G^*)$ and $(\hat \theta, \hat G)$ can be bounded using the following data processing inequality:
\begin{equation}\label{eq:data-processing-ineq-2}
    I(\theta^*, G^*; \hat \theta, \hat G) \leq I(\theta^*, G^*; X^*)\,.
\end{equation}
In the hierarchical context, $I(\theta^*, G^*; X^*)$ sets the reconstruction limit.
By the chain rule, we have
\begin{equation}
    I(\theta^*, G^*; X^*) = I(G^*;X^*) - I(\theta^*; X^*|G^*)\,,
\end{equation}
for which the second term of the RHS is zero, by the conditional independence of $X^*$ and $\theta^*$ given $G^*$.
We are left with the mutual information upper bound $I(\theta^*, G^*; X^*) = I(G^*;X^*)$, which is equal to the non-hierarchical case.
This means that the reconstruction limit is always set by $I(G^*;X^*)$, even if the hyperparameters $\theta^*$ are not marginalized over.

\section{Data-driven reconstructability and model selection}
\label{sec:data-driven-reconstructability}
Until now, we have assumed that the TDG model $M^*$ was known to compute the mutual information $I(G^*;X^*)$.
Outside of theoretical settings however, the TDG process is typically unknown.
Hence, we generally cannot evaluate the true reconstruction limit, although we may have access to many realizations of $X^*$ which should help get closer to it.
Three remarks are in order.
First, the reconstructability $\Psi^*$ is independent of the observations; it strictly depends on $M^*$.
Second, any generative model $M$
has a reconstructability, which is calculated identically to Eq.~\eqref{eq:reconstructability}.
In other words, the condition that $M$ is capable of generating new data is crucial to our ability to calculate a reconstructability value.
However and thirdly, the reconstructability of $M$ differ in two ways from $\Psi^*$ related to the actual reconstruction limit of the data: \emph{(i)} their values are potentially different and \emph{(ii)} the data generation process is $M^*$, not $M$.
Consequently, we can leverage the reconstructability of $M$, with these considerations in mind, to get a data-driven proxy of the true upper bound $\Psi^*$.

\subsection{Reconstruction index based on information gain}
\label{subsec:recon_index}
To bring back the dependency of the reconstructability on the observations, we take a similar approach as before and start with an information measure.
For a model $M$ and any instance $x\in\mathcal{X}$, the data-driven version of mutual information is called the \emph{information gain}~\cite{mitchell1997machine}, and it is defined as
\begin{align}\label{eq:infogain-1}
    \mathcal{I}_M(x) 
    &= -\expect[G|X=x]{\log \left(\frac{P(G|X)}{P(G)}\right)}\notag\,, \\ 
    &= \sum_{g\in\mathcal{G}}P(G=g|X=x)\log \left(\frac{P(G=g|X=x)}{P(G=g)}\right)\,.
\end{align}
Note that the expectation of the information gain yields back the mutual information between $G$ and $X$, i.e., $\expect[X]{\mathcal{I}_M(X)} = I(G;X)$.
The information gain measures the reduction in the entropy of a variable $G$ achieved by learning the state $x$ of another variable $X$.
It is primarily used in feature selection, especially decision tree training, where it is used as a criterion for how to best split the data~\cite[Chapter 3]{mitchell1997machine}.  
Like the mutual information, the information gain can be shown to be non-negative (see Appendix~\ref{app:information-gain}) and upper-bounded:
\begin{equation*}
  0 \leq \mathcal{I}_M \leq \Lambda_M\,,
\end{equation*}
where
\begin{equation}\label{eq:reconstruction-index-norm}
  \Lambda_M(x) = -\expect[G|X=x]{\log P(G)}\,
\end{equation}
is the maximum value of the information gain, and can be interpreted as the cross-entropy between the reconstruction posterior and the prior probabilities of $M$.
It is therefore convenient to define a normalized version of the information gain, which we refer to as the \emph{reconstruction index}:
\begin{equation}\label{eq:reconstruction-index}
  \psi_M = \frac{\mathcal{I}_M}{\Lambda_M}\,.
\end{equation}

Like the reconstructability, the reconstruction index $\psi_M$ is bounded between $0$ and $1$.
However, it differs mainly in that $\psi_M$ may yield different values for different datasets $x$.
In addition, there is a subtle difference in their interpretations, as we will see in the following sections.
Indeed, the information gain, on which the reconstruction index is based, is the Kullback-Leibler (KL) divergence between the posterior and the prior of the reconstruction model.
As a result, it quantifies how different the posterior of the reconstruction model is from the prior.
When $\psi_M = 0$, the posterior and the prior are identical---no information is gained from knowing the data.
On the other hand, when $\psi_M = 1$, the posterior probability mass is entirely located on a single graph, which is reflected in the fact that the KL divergence is maximized.

\subsection{Interpretation of the reconstruction index under incorrect assumptions}
\label{sec:underfitting}
\begin{figure}
  \centering
  \includegraphics[scale=0.9]{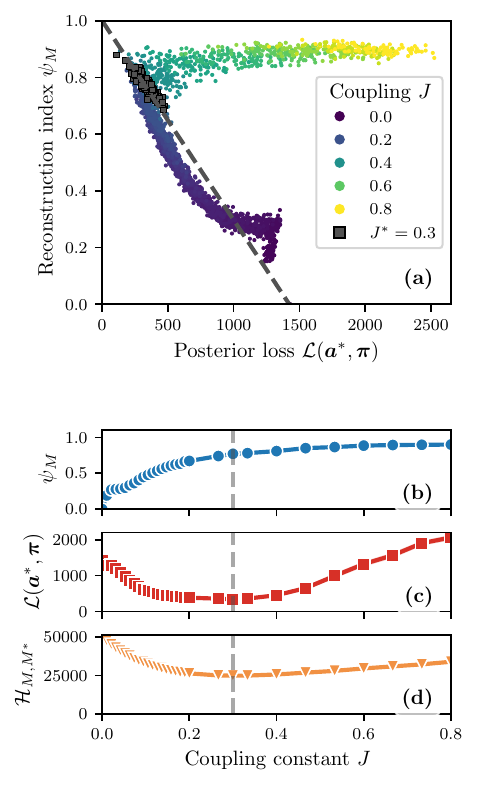}
  \caption{Effect of varying the coupling constant on the validity of the reconstruction index. 
  We generated time series of the Glauber dynamics with fixed $J^*=0.3$ on Erd\H{o}s-R\'enyi graphs with $N=100$ nodes and $E=250$ edges, then reconstructed the graphs using the same Glauber model with other coupling constants $J$, used during the inference.
  Panel (a) shows the relationship between the reconstruction index $\psi_M$ and the posterior loss $\mathcal{L}(\vec{a}^*, \vec\pi)$ between the true graphs and the posterior---each point corresponding to a different realization of the TDG process (graph and observations) from which we reconstructed the graph. 
  Panels (b--d) respectively show the reconstruction index $\psi_M$, posterior loss $\mathcal{L}(\vec{a}^*, \vec\pi)$, and evidence cross-entropy $\mathcal{H}_{M, M^*}$
  (Eq.~\eqref{eq:evidence-cross-entropy}) as functions of $J$. The dashed vertical line shows where $J=J^*$.
  We color-coded the points according to $J$, as shown in the legend, including the true value $J^*$ (grey squares). 
  As in Fig.~\ref{fig:recon-vs-metrics}, we show the linear relationship between the reconstructability and the posterior loss (Eq.~\eqref{eq:linear-relationship-u-celoss}) with the dashed line in (d). 
  Glauber time series were generated with $T=500$ time steps, and we generated 24 realizations with random initial conditions for each value of $J$ between 0 and 0.8 (like in Fig.~\ref{fig:recon-vs-metrics}, we show only a few values in the legend of (d)).
  In panels (b--d), we show the 90\% confident intervals around the mean (displayed by the markers), although they are too small to be visible.
  }
  \label{fig:recon-convergence}
\end{figure}

We must be careful in our interpretation of the reconstruction index, as its value can be misleading if not used in the correct way.
Figure~\ref{fig:recon-convergence} shows how the reconstruction index behaves when the reconstruction model is incorrect to different extents.
In this example, we generated time series of the Glauber dynamics with a given coupling constant $J^*$, and reconstructed the graphs using the same Glauber model, but typically with an erroneous coupling constant $J \neq J^*$.
As we can see in Fig.~\ref{fig:recon-convergence}(a), the reconstruction index keeps increasing as $J$ gets larger, even when it gets larger than $J^*$.
While the reconstruction index is larger for $J>J^*$, the posterior loss actually shows, as expected, that the reconstruction becomes worse.
Figure~\ref{fig:recon-convergence}(a) also illustrates how the reconstruction index correlates with the posterior loss, depending on $J$.
Indeed, as $J$ gets closer to $J^*$, the reconstruction index converges to the true reconstructability of the reconstruction model, which increasingly becomes linearly related to the posterior loss as previously shown 
(see the Appendix~\ref{app:posterior-loss})
In this regime, the reconstruction index is a good proxy of the true reconstructability because $M$ properly approximates the behavior of $M^*$.

The behavior of the reconstruction index when the reconstruction model is incorrect raises some important remarks.
Recall that, fundamentally, the reconstruction index is a normalized version of the KL divergence between the posterior and the prior of the reconstruction model.
Therefore, it is perhaps not surprising that we lose the correspondence between reconstruction index and performance established in Sec.~\ref{sec:reconstructability-performance} when the model is incorrect.
Indeed, having a posterior that is very different from the prior implies a high reconstruction index even though the posterior distribution is actually wrong.

Maximizing the reconstruction index can also lead to inadequate modeling of the observed data.
Consider the following alternative but equivalent form of the information gain:
\begin{equation}\label{eq:infogain-2}
  \mathcal{I}_M(x) =\,\, \expect[G|X=x]{\log P(X|G)} - \log P(X=x)\,.
\end{equation}
In this formulation, the two terms---i.e., the expected log-predictive and the log-evidence, respectively---are in opposition.
The first term is maximized when the model is good at describing the data using the posterior graphs, while the second is maximized when the model describes the data correctly altogether.
The log-evidence is even used as a measure of goodness-of-fit for model selection, as we will see in the next section.
Yet, maximizing the information gain is equivalent to maximizing the expected log-predictive and minimizing the log-evidence, which is why incorrect models may be selected by this criterion.
Following these remarks, we devote the next section to describing a principled approach to adequately interpret the reconstruction index and use it in the context of data-driven reconstruction.

\subsection{Role of evidence-based model selection}
\label{sub:evidence-model-selection}

Maximizing evidence as a criterion for model selection is a well-known practice in Bayesian modeling~\cite{kass1995bayes}.
In particular, Bayes factors are ratios between the evidence of two models, say $M_1=(G_1, X_1)$ and $M_2=(G_2, X_2)$:
\begin{equation}
  B_{M_1, M_2}(x) = \frac{\zeta_1(x)}{\zeta_2(x)}\,, 
\end{equation}
where $\zeta_{M_i}(x) = P(X_i=x)$ is the evidence of model $M_i$ for $x$.
If $B_{M_1, M_2}(x) > 1$, $M_1$ is better supported by the data $x$  than $M_2$.
From an information-theoretic perspective, the minimization of the \emph{evidence cross-entropy} (CE)---which is equivalent to maximizing the evidence---can be shown to be a necessary condition for finding the TDG model, whose evidence function is $\zeta_{M^*}(x)$.
Indeed, for a reconstruction model $M$ with evidence function $\zeta_M(x)$, the evidence CE is expressed as
\begin{equation}\label{eq:evidence-cross-entropy}
\begin{split}
  \mathcal{H}_{M^*, M} 
  &= -\expect[X^*]{\log \zeta_M(X^*)}\,,\\
  &= -\sum_{x^*\in\mathcal X} \zeta_{M^*}(x^*) \log \zeta_M(x^*)\,.
\end{split}
\end{equation}
Equation~\eqref{eq:evidence-cross-entropy} is minimized when $X^*$ and $X$ are equal in distribution~\cite{cover2006elements}.
Note that it is a necessary condition to find the correct TDG model, but may not be a sufficient one, as it is easy to show that, in the general case, many reconstruction model may have the same evidence distribution, but different posterior distributions.

The problem of evidence-based model selection is the computation of the evidence itself which is often intractable in practice.
In fact, this is the case for most graph models, where the evaluation of the evidence requires graph enumeration.
This problem also arises in the evaluation of both the information gain and the mutual information.
Fortunately, the same numerical techniques can be used to evaluate the evidence and the information gain simultaneously, as the two are related to each other.
In Ref.~\cite{murphy2024duality}, we showed that variational mean-field methods provide efficient approximations for both the mutual information and the evidence. 
The same techniques are used here (see Appendix~\ref{app:numerical-techniques}).

Model selection is crucial for the validity of the reconstruction index as a proxy of reconstruction performance.
Suppose we have many observations $(x_1, x_2, ...)$ and used the reconstruction model $M^*$.
The empirical average of the information gain of some model $M$ converges to $\expect[X^*]{\mathcal{I}_M(X^*)}$.
Now, assume for a moment that $M$ in fact maximizes the expected log-evidence.
This implies, as we mentioned before, that $X^*$ and $X$ are equal in distribution.
In turn, the empirical average of information gain becomes equal to the mutual information for model $M$, i.e., $I(X;G)$, which we recall is the reconstruction limit of $M$.
This means that empirical average of information gain converges to the reconstructability and that ultimately the reconstruction index naturally extends the concept of reconstruction limit to real systems observed only through data.

When the reconstruction model does not minimize the evidence CE, the picture becomes more nuanced, as shown in Fig.~\ref{fig:recon-convergence}.
As the evidence CE decreases, the correlation between the reconstruction index and the posterior loss increases.
Our ability to identify the reconstruction limit without knowing the true model or network structure is therefore as good as the reconstruction model's ability to describe the data.
This key conceptual observation leads us to conclude that we can indeed leverage the reconstruction index as a proxy for assessing the reconstructability of real networks, provided it is interpreted in conjunction with the posterior loss, as in Fig.~\ref{fig:recon-convergence}(a).

\section{Network reconstructability in empirical networks}
\label{sec:reconstructability-real-systems}
\begin{figure*}
  \centering
  \includegraphics[width=1\textwidth]{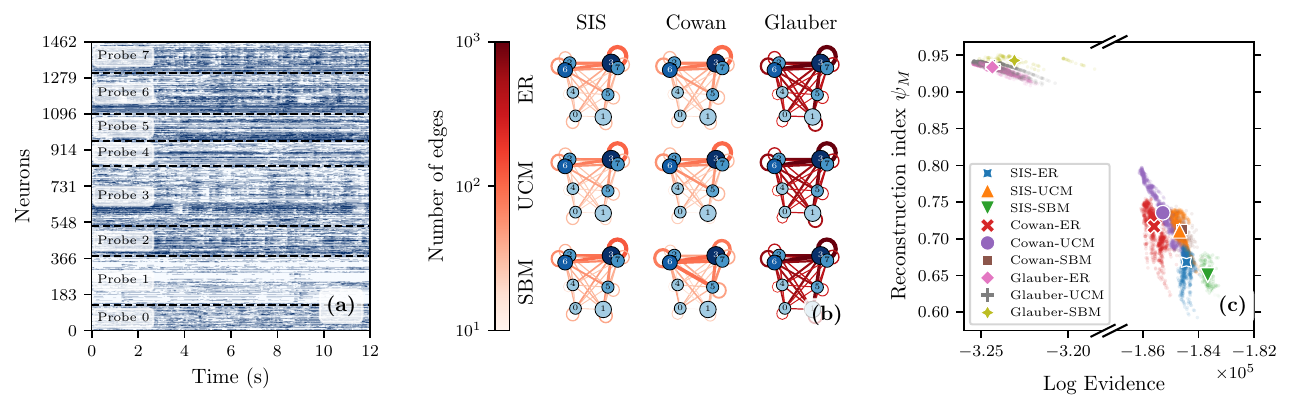}
  \caption{Reconstruction from spontaneous neuronal activity in the mouse brain~\cite{steinmetz2019eight,stringer2019spontaneous}:
  (a) Raster plot of the 1462 monitored neurons, (b) reconstruction of the probe network using different reconstruction models and (c) reconstructability diagram. 
  In panel (a), the neurons are ordered by the probe they were measured from. 
  Each spike is represented in blue. 
  Panel (b) shows the posterior average network projected onto the probes, as predicted by each reconstruction model where rows correspond to different graph models (see Appendix~\ref{app:graph-models}), and columns to different dynamics models (see Table~\ref{tab:dynamics}).
  The color of an edge connecting two probes shows the absolute number of edges and the thickness indicates the average proportion among all the edges. 
  The size of the probe nodes is proportional to the number of neurons monitored by the probe, and the color indicates the measured number of spikes. 
  The node locations correspond to the actual probe locations in the mouse brain obtained from~\cite{steinmetz2019eight}. 
  The reconstruction index as a function of the model log evidence is shown in panel (c), comparing the different models.
  Small markers are estimated by a single Markov chain and large markers are the average of these estimations.
  In these experiments, the parameters of the graph prior and likelihood are inferred jointly with the graph.
  For additional details about the inference procedure, we refer to Appendix~\ref{app:inference-brain}.
  }
  \label{fig:spiking-neurons}
\end{figure*}

Reconstructing empirical graphs represents a technical and conceptual challenge.
The true network structure being unknown, it is hard to quantify how close the predicted graph is to the true one, let alone calculate its actual reconstructability.
In light of our exploration in Sec.~\ref{sec:data-driven-reconstructability}, however, we have shown that the reconstruction index $\psi_M$ can offer a means to approximate the reconstructability, if certain conditions are met.
Additionally, our analysis shows that a correctly calibrated reconstruction index $\psi_M$ predicts the error in a reconstructed network in comparison with the true one.
Under these considerations, we present a principled method based on the reconstruction index to assess the validity of network reconstruction.

In this procedure, we assume $x$ to be some time series generated by a hidden process $M^*$, from which we wish to infer a network.
Next, the procedure goes as follows:
\begin{enumerate}
    \item Select a set of $d$ reconstruction model candidates $\mathcal{M} = \{M_1, M_2, ..., M_d\}$.
    \item For each candidate $M$, sample a reconstructed graph ensemble $\hat{\mathcal{G}}_M$ using the posterior of $M$.
    \item Calculate for each candidate $M$ the evidence $\zeta_M(x)$ and the reconstruction index $\psi_M(x)$ using $\hat{\mathcal{G}}_M$ (see Sec.~\ref{app:numerical-techniques} for detail).
    \item Choose the reconstruction index $\psi_{\hat{M}}(x)$ of the model $\hat{M}$ with the highest evidence:
    \begin{equation}
        \hat{M} = \argmax_{M\in\mathcal{M}} \zeta_M(x)\,.
    \end{equation}
    The index $\psi_{\hat{M}}(x)$ is the output of the procedure.
\end{enumerate}
Some remarks about this method are in order.
Indeed, as shown in Sec.~\ref{sub:evidence-model-selection}, the validity of the reconstruction index heavily relies on the model's aptitude to represent the data.
Hence, it is paramount that an adequate set of model candidates is selected at first.
These models may differ in their underlying assumptions, via their prior, hyper prior, and/or likelihood functions, as previously stated.
If $\mathcal{M}$ contains a model that resembles the TDG process, this procedure will generate a reconstruction index $\psi_{\hat{M}}(x)$ that closely approximates the true reconstructability of the process $\Psi^*$.

Of course, determining if such a model is in $\mathcal{M}$ is hardly feasible experimentally.
We work around this issue by performing a validation after the procedure via a predictive posterior check of $\hat{M}$.
This validation proceeds as follows:
\begin{enumerate}
    \item Generate a sample $\set{\hat{x}_1, ..., \hat{x}_K}$ of synthetic data with $\hat{M}$, assuming its parameters $(\theta, \phi)$ are sampled from the model's posterior given $x$;
    \item Calculate some test quantities $\tau(\hat{x}_k)$, i.e., statistics used for comparison, for each sample $\hat{x}_k$, generating a set of samples $\mathcal{T} = \set{\tau(\hat{x}_1), ... , \tau(\hat{x}_K)}$;
    \item Compare $\tau(x)$ with $\mathcal{T}$.
\end{enumerate}
If $\tau(x)$ is typical in $\mathcal{T}$, then we can be sure $\hat{M}$ is statistically similar to $M^*$.

In what follows, we will demonstrate this method in two different empirical use cases. 
The first use case reflects a realistic modeling scenario, where only the time series on which reconstruction is performed are known.
From a theoretical perspective, this use case presents challenges that falls outside of the scope of this paper (we will discuss those in the following section).
The second use case lift those limitations by reconstructing real networks from synthetic time series.

\subsection{Reconstruction from empirical neuronal spiking data}
\label{sec:reconstructability-brain-networks}
We consider the spontaneous activity of 1462 neurons from the dorsal cerebral cortex of a mouse recorded over 20 minutes using eight Neuropixel probes [Fig.~\ref{fig:spiking-neurons}(a)]~\cite{stringer2019spontaneous}.
Starting from the recorded spike times of the neurons, indicating when they fire, we create a binary time series of the activity of each neuron.
In these binary time series, a `1' marks the moments when a neuron is fired, and a `0' when it is not. 
For more detail regarding our data processing procedure, see Appendix~\ref{app:inference-brain}.
Following our method described at the beginning of Sec.~\ref{sec:reconstructability-real-systems}, we infer the network using a variety of Bayesian reconstruction models.
As a result, we get the reconstructed networks shown in Fig.~\ref{fig:spiking-neurons}(b), where we aggregated the edges connecting the neurons of all pairs of probes, thus illustrating how they interact.
Finally, we compute the log evidence for each model and select the model with the highest one (see Appendix~\ref{app:numerical-techniques} for details on the evidence estimation). 
In doing so, we extract the reconstruction index that is, in principle, closest to the reconstruction limit, given the set of considered models.
Figure~\ref{fig:spiking-neurons}(c) shows a diagram of both measures for each model.

In our example, the analyzed reconstruction models are combinations of time series likelihoods---SIS, Cowan and Glauber---and graph priors---the ER model, the configuration model with a uniform degree sequence hyperprior (UCM) and the stochastic block model (SBM).
The ER model, having a uniform distribution, is the most entropic model, followed by the SBM and the CM.
Additional details about the graph prior and about these reconstruction models are given in Appendices~\ref{app:graph-models} and~\ref{app:markov-chain}, respectively.
Of course, these models oversimplify the observed neuronal activity.
Moreover, certain critical factors are not captured in the current dataset, such as the latency and deactivation rates of neuronal activity, as well as the substantial number of neurons undetected by the probes, which could contribute as input currents to the modeled neurons. 
In addition, the lack of detailed connectomic data for such small brain regions prevents a rigorous assessment of the accuracy reconstructed probe network.
The purpose of this analysis is therefore not to perform the most accurate reconstruction but to illustrate the complete procedure as well as the results it generates.

Among the considered models, the SIS model with a SBM prior is the one achieving the highest log evidence. 
As detailed in Appendix~\ref{app:inference-brain}, the posterior predictive checks of the inferred SIS model suggests that its reconstruction index of approximately $67\%$ is reasonable.
Additionally, given that the network contains $1462$ neurons and that the estimated number of edges for this model is approximately $1722$ (see Table~\ref{tab:edge_count_stats}), the average degree is $2.35$, indicating a network far sparser than anatomical data would predict. Indeed, a recent connectomic study of a subvolume of the mouse visual cortex~\cite{turner2022reconstruction}---part of the dorsal cortex studied in~\cite{stringer2019spontaneous}---reports a connection probability of $0.054$. If we hypothetically assume a similar probability across the entire dorsal cortex, the expected total number of edges would be approximately $1.15\times 10^{5}$, corresponding to an average degree of about $79$.
This suggests that, although our estimation of reconstructability is not close to zero, the inferred network does not seem to account for most of the neural activity.
Of course, considering the decreasing tendency observed in Fig.~\ref{fig:spiking-neurons}(c), more detailed neuronal models---better suited for these data and with potentially higher log evidence---could yield reconstruction indices even lower than $67\%$, with possibly denser inferred networks.
In the following section, we circumvent the limitations of the neuronal activity data by transitioning to a controlled setting, where synthetic activity data is used to reconstruct empirical networks.

\subsection{Reconstruction of empirical graphs from synthetic activity data}
While reconstructing empirical graphs, we assume that all observations come from the same graph $g^*$.
Hence, the graph prior, whose associated random variable is $G$, plays an important role: Injecting prior information about $g^*$.
Depending on the value of $P(G=g^*)$, the graph prior may either improve the reconstruction or impede it.
Consequently, the amount of information that one may need to achieve a desired level of reconstruction accuracy may change depending on the choice of $P(G)$.

\begin{table}[]
    \centering
    \begin{tabular}{l|c c c c}
        \hline \hline
         Graph &  ER & UCM & CM & SBM\\ \hline
         Karate club & 343.69 & 316.52 & 200.82 & 328.61\\ 
         Political books &  2267.99 & 2177.04 & 1756.70 & 2158.49\\ 
      \hline \hline
    \end{tabular}
    \caption{Negative log probability---i.e., $-\log P(G=g^*)$---of the graphs considered in Fig.~\ref{fig:empirical-graphs} using the Erd\H{o}s-Rényi (ER) model, the configuration model with uniform degree sequence prior (UCM), the configuration model with given degree sequence (CM) and the stochastic block model (SBM).}
    \label{tab:graph-prior-values}
\end{table}

\begin{figure*}
  \centering
  \includegraphics[width=0.9\textwidth]{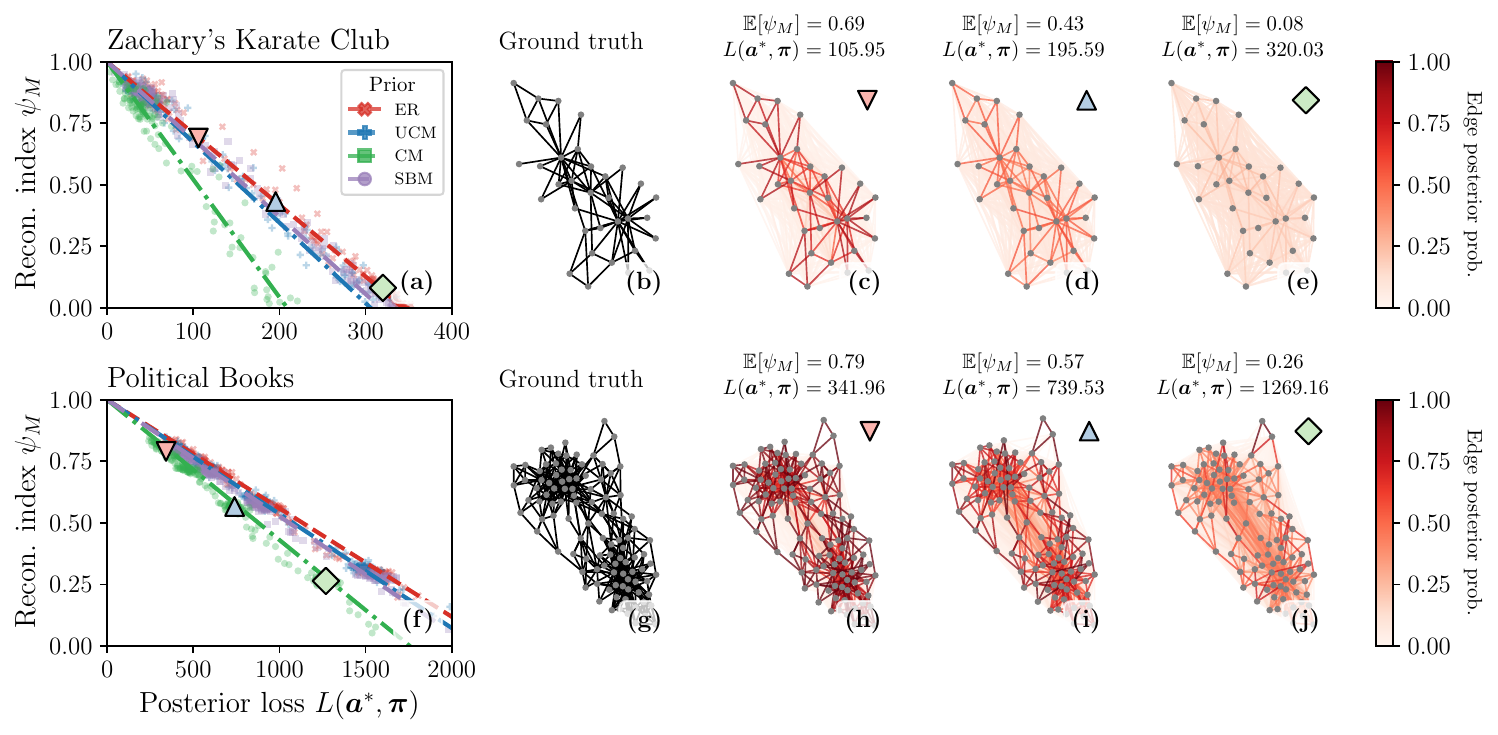}
  \caption{ 
    Reconstruction indices of empirical graphs with different graph prior models:  (top) SIS dynamics on the Zachary's karate club~\cite{zachary1977information} and (bottom) Voter dynamics on the Political books network. 
    We show in panels (a) and (f) the reconstruction indices $\psi_M$ as a function of the posterior loss. 
    We consider different values of dynamics parameters to populate the diagrams: for Zachary's karate club we fixed the infection probability to $\lambda\in\set{0.1, 0.12, 0.15, 0.2, 0.3}$, and for the Political books network, we let $\alpha_0\in\set{0.001, 0.01, 0.1, 0.25, 0.5}$---we omit illustrating their values in the plots for simplicity. 
    We use and fix these parameter values within the model during the inference.
    For each combination of graph model and dynamics parameters, we generated 48 time series of $T=300$ steps and performed reconstruction of each of them individually. 
    Each point in (a) and (f) corresponds the reconstruction index and posterior loss of one of these time series.
    In each plot, the different symbols and colors indicates the graph prior model used for the reconstruction: The Erd\H{o}s-R\'enyi model (ER, blue diagonal crosses), the configuration model with uniform degree sequence prior (UCM, orange crosses) and with the correct degree sequence (CM, red circles), and the stochastic block model (SBM, green squares). 
    The lines correspond to the scaling of $\psi_M$ with respect to the posterior loss [Eq.~\eqref{eq:reconstruction-index}]. 
    In panels (b--e) and (g--j), we show the true network $g^*$ (far left) followed by the reconstructed graphs, as illustrated by their respective posteriors, of three different models. 
    We indicate on top of each example the corresponding expected reconstruction index and posterior loss, and we highlight their location in the diagrams of (a) and (f) using the symbols (inverted triangle, triangle and diamond). 
    For panels (c--e), we choose the posteriors of the ER model such that (c) $\lambda=0.1$, (d) $\lambda=0.15$ and (e) $\lambda=0.2$. 
    For panels (h--j), we choose the posteriors of the CM where (h) $\alpha_0=0.5$, (i) $\alpha_0=0.25$ and (j) $\alpha_0=0.1$.
    }
  \label{fig:empirical-graphs}
\end{figure*}

Our method can also be used to assess the role of the prior, when using empirical graphs.
However, since we consider here synthetic data generation process, we can omit the posterior predictive checks.
Figures~\ref{fig:empirical-graphs}(a) and (f) show the reconstruction index of two empirical graphs, namely Zachary' karate club~\cite{zachary1977information} and the Political books network~\cite{kreb2004political}, as a function of the posterior loss for different graph prior models.
Here, the reconstruction indices $\psi_M$ are calculated using Eq.~\eqref{eq:reconstruction-index} like before, where all $x^*$ are generated using the same graph $g^*$.
For this analysis, in addition to the previous graph priors, we consider the standard CM where the degree sequence is given.
The standard CM is less entropic than the UCM since it is given all the information of the degree sequence---it does not need to be inferred from $X$ as for the UCM.
The graph support of the CM is also quite smaller than that of the other priors, which should improve the reconstruction.
We generated many synthetic observations on these two graphs $g^*$ to perform the reconstruction: A spreading epidemics on the karate club social network using the SIS model, and the Voter model for the Political books network, as to simulate the propagation of political opinions.
For both examples, we chose many values of parameters for the dynamics to present the complete range of reconstructability scenarios. 
To highlight the role of the graph prior, we reconstruct all graphs with the TDG data models.
This means that, in this case, $\psi_M$ is indeed a point-wise measure of reconstructability.

The reconstruction index is shown to scale linearly with the error, as measured by the posterior loss, identically to Fig.~\ref{fig:recon-convergence}(a).
This shows that, even though only one graph is used to generate the data, the relationship between $\psi_M$ and performance still holds.
This is also observed in Figs.~\ref{fig:empirical-graphs}(b--e) and (g--j), where examples of reconstructed graphs with increasing reconstruction indices are shown with their corresponding---and increasingly accurate---posterior. 

Moreover, notice how the reconstruction index scaling changes as a function of the prior, where the slope is precisely given by $\Lambda_M^{-1}$ [Eq.~\eqref{eq:reconstruction-index-norm}].
In both cases, the graph model with the steepest slope is the CM, which also has the most prior information about $g^*$ as shown by the prior negative log probability in Table~\ref{tab:graph-prior-values}.
This happens because $\Lambda_M$ and the prior probability $P(G=g^*)$ are intrinsically related to one another, since $\Lambda_M$ is the posterior average of the prior probability.
Hence, as the posterior becomes more concentrated around $g^*$, $\Lambda_M$ converges to $P(G=g^*)$.
The relationship between the $\psi_M$ scaling and the prior implies that $\psi_M$ tends to diminish faster as a function of the posterior loss, as the prior gets more informative.
In other words, for two models with identical reconstruction indices, the one with a graph prior more concentrated around $g^*$ generates a more accurate reconstruction.

The reconstruction limit, as measured by $\psi_M$, changes as a function of the graph prior: If information about $g^*$ is \emph{a priori} given, this same information cannot be reconstructed from realizations of $X^*$. 
Thus, it is no longer taken into account in $\psi_M$ which by construction factors out the contributions of the prior.
This intuition can be mathematically studied if we let the prior put more and more weight on $g^*$.
In Appendix~\ref{app:reconstructability-delta}, we prove that as the graph generative model converges to a Kronecker delta distribution, i.e., 
\begin{equation}
    P(G^*=g) = \begin{cases}
        1 & \text{if $g = g^*$}\\
        0 &\text{otherwise}
    \end{cases}\,,
\end{equation}
the reconstructability $\Psi^*$ converges to zero, except if the mutual information is maximized in which case it is always equal to 1.

\section{Conclusion}
\label{sec:conclusion}
To what extent is a complex network reconstructible~?
Our ability to reconstruct is strongly constrained by the information content of the underlying structure within the data, making perfect reconstruction generally infeasible.
The best reconstruction  therefore amounts to finding a network that reaches a reconstruction limit, extracting all available information in such a way that no further improvement can be achieved on average.

Our information-theoretic framework characterizes this network reconstruction limit, which is closely tied to the true generating process of the observed data.
The reconstruction limit is analogous to the detectability limit in community detection~\cite{decelle2011inference, ghasemian2016detectability, young2017finite} in that it is algorithm independent.
We find that this limit is expressed in terms of the reconstructability---a normalized version of the mutual information between the graph and the data of the true data generating process.
While a small reconstructability implies a bad performance regardless of the reconstruction model, a high reconstructability implies that good performance can be achieved with the appropriate model.

Our approach is general and can be extended to real modeling settings, where the data is limited and the reconstruction model is unknown.
Using the same principles, we defined the reconstruction index, analogous to the reconstructability, that is also data-dependent and can be used as a proxy of the reconstruction performance.
When coupled with evidence-based model selection, the reconstruction index is an appropriate performance measure, even when the graph is unknown.
This further emphasizes the importance of model quality in network reconstruction~\cite{peel2022statistical}.

Finally, we presented different applications of our framework using real networks and real time series data.
We showed how to use the reconstruction index on spiking neural networks.
Our analysis suggests that the reconstructability of the network formed by the recorded neurons in the mouse brain is approximately 67\%, which is consistent with experimental studies of brain networks reporting structure-function couplings of about 40\% in humans \cite{baum2020development} and up to 68\% in smaller animal models, such as zebrafish~\cite{legare2024structural}.
We believe that more work is needed in the inference process, primarily because of the simplicity of the reconstruction models used in this analysis and the incompleteness of the data.
A more thorough analysis of such a case study could reveal that certain neuronal activity datasets are insufficient to build a comprehensive picture of the functional activity of the brain.
Additionally, although the reconstructability is based on random variables and ensembles, we demonstrate that our framework can be reliably used on single instances of graphs.
In this context, the reconstruction index can be used to determine the reconstruction limit, even if the true graph is unknown, as it is shown to correlate strongly with the error between the inferred graphs and the true one.

We envision a future where network reconstruction applications incorporate a reconstructability analysis in their pipeline, such as the one presented in Sec.~\ref{sec:reconstructability-real-systems}.
By doing so, the reconstruction index would indicate how informative the reconstructed networks are and perhaps inform us on how they should be used within the said applications.
Of course, there is still plenty of work to be done on this front, such as improving the computational methods required to compute the reconstruction index as they do not scale well to large networks, and improving the reconstruction models themselves as we have alluded to earlier.
Some of these models might also require modifying our framework, for example in the case of weighted and directed networks.
These specific models could prove considerably valuable for the neuroscience community and, more broadly, for complex systems research.

\appendix

\section{Graph priors}
\label{app:graph-models}

In the paper, we use different random graph models as graph priors for Bayesian network reconstruction.
These models are undirected and unweighted and may include self-loops and multiedges, although our general framework is not restricted to these assumptions.
Indeed, one could consider directed or weighted graphs as well; as long as the set of possible graphs remains countable.
We use the adjacency matrix, denoted $\vec{a}$, in order to define the probability distribution of some of these models, where $a_{ij}$ counts the number of edges connecting nodes $i$ and $j$.
To simplify the notation, we will sometimes express a graph $g$ directly with its adjacency matrix $g=\vec{a}$.
We use the convention that $a_{ii}$ is always a multiple of 2.
Below, we describe these priors in more detail.

\subsection{Erd\H{o}s-Rényi model}
The Erd\H{o}s-Rényi (ER) model corresponds to the maximum entropy random graph model, i.e., the uniform distribution over all simple graphs with $N$ nodes and $E$ edges, such that
\begin{equation}
    P(G|E=e) = \binom{\frac{N(N-1)}{2}}{e}^{-1}\,,
\end{equation}
where we recall that $\binom{n}{k}$ is the binomial coefficient. 
The ER model is also generalizable to loopy multigraphs, where
\begin{equation}
    P(G|E=e) = \mutilsetcoeff{\frac{N(N + 1)}{2}}{e}^{-1}\,,
\end{equation}
such that $\mutilsetcoeff{n}{k} = \binom{n + k - 1}{k}$ counts the number of possible multisets of size $k$ composed of $n$ different objects---i.e., multiedges.

Note that the number of edges $E$ must be provided to the ER model, and in the other graph models described below.
This means that $\theta=E$ is the hyperparameter of ER graph prior and that $E$ should be inferred.
We use the prior $P(E)$ to weigh in the number of edges.
In most of our experiments, the number of edges is fixed to a specific value $e^*$, meaning that $P(E=e) = \delta(e, e^*)$, where $\delta(m, n)$ is the Kronecker delta function.
However, in Sec.~\ref{sec:reconstructability-brain-networks}, as $E$ is unknown in this case, we use a geometric prior of the form
\begin{equation}
    P(E=e) =  \frac{\bar\lambda^e}{(\bar{\lambda} + 1)^{e + 1}}\,,
\end{equation}
where $\bar{\lambda}$ is a parameter that fixes the expected number of edges.
See also Appendix~\ref{app:inference-brain} for further detail about the complete inference procedure.

\subsection{Configuration model}
The configuration model (CM) describes an ensemble of loopy multigraphs where the degree sequence is given~\cite{fosdick2018configuring}.
From a network reconstruction perspective, the CM can also be used as a prior, assuming that its probability factors as follows
\begin{equation}
    P(G, \vec{k}, E) = P(G|\vec{k}) P(\vec{k}| E) P(E)\,,
\end{equation}
where $P(G|\vec{k})$ is the graph likelihood given the degree sequence $\vec{k}$, $P(\vec{k}|E)$ is the prior over the degree sequence and $P(E)$ is again the prior over the number of edges (same as in the ER model).
In the CM, half-edges (or stubs) are considered distinguishable and a realization of the model is generated by randomly pairing all available half-edges.
Hence, the probability of generating pairings leading to a graph $g$, whose adjacency matrix is $\vec{a}$, given its degree sequence $\vec{\kappa}$ is
\begin{equation}
    P(G=\vec{a}|\vec{k}=\vec{\kappa}) = \frac{\prod_{i=1}^N \kappa_i!}{\prod_{i<j}a_{ij}! \prod_{i=1}^N a_{ii}!!}\,,
\end{equation}
where $(2n)!! = 2^{n} n!$ is the double factorial of $2n$, i.e., the product of all even numbers up to $2n$.

Furthermore, only one degree sequence, denoted $\vec{\kappa}^*$, is considered in the standard formulation of the CM. 
This results in a delta degree sequence of the form
\begin{equation}
    P(\vec{k}=\vec{\kappa}) = \delta(\vec{\kappa}, \vec{\kappa}^*)\,.
\end{equation}
When the degree sequence is unknown, we use the uniform non-informative prior
\begin{equation}
    P(\vec{k}|E=e) = \mutilsetcoeff{N}{2e}^{-1}\,,
\end{equation}
where $\mutilsetcoeff{N}{2e}$ counts the number of possible degree sequences for a graph of $N$ nodes and $E$ edges.

\subsection{Stochastic block model}
The stochastic block model (SBM), in its microcanonical version~\cite{peixoto2017nonparametric}, closely resembles the ER model, where edges are picked uniformly at random. 
However, unlike the ER model, each node $i$ is associated with a random block $b_i\in\{1, 2,\dots, B\}$ and instead the number of edges $e_{rs} = \sum_{ij} a_{ij} \delta(b_i, r) \delta(b_j, s)$ connecting two blocks $r$ and $s$ is fixed such that there are $E$ edges in total.
We summarize the node partition as a tuple $\vec{b}=(b_i)_{i=1..N}$ and the number of edges between blocks by the edge matrix $\vec{e} = (e_{rs})_{r,s=1..B}$.
The blocks are required to be non-empty.
Since $B$ is the number of non-empty blocks in $\vec{b}$ and $\vec{e}$ are completely determined by the graph and $\vec{b}$, $\theta = (\vec{b}, E)$ are the hyperparameters of the SBM, then $\vec{b}$ and $E$ must be inferred jointly with the graph.
In theory, one could factor the joint prior probability $P(G, E, \vec{b})$ as $P(G|E, \vec{b}) P(E) P(\vec{b})$ assuming $E$ and $\vec{b}$ are independent.
However, it is more convenient to factor the prior also using $\vec{e}$ and $B$ as intermediate random variables, following
\begin{align}
    P(G, E, \vec{b}) &= P(G, E, \vec{e}, \vec{b}, B) \nonumber \\
    &= P(G|\vec{e}, \vec{b}) P(\vec{e}|\vec{b}, E)  P(E) P(\vec{b}|B) P(B)\,,
\end{align}
where
\begin{equation}
    P(G|\vec{e}=\vec{\epsilon}, \vec{b}=\vec{\beta}) = \prod_{r<s} \binom{n_r n_s}{\epsilon_{rs}}^{-1} \prod_r \binom{\frac{n_r (n_r + 1)}{2}}{\frac{\epsilon_{rr}}{2}}^{-1}\,
\end{equation}
and $n_r = \sum_{i=1}^N \delta(\beta_i, r) $ counts the number of nodes in block $r$ for the partition $\vec{\beta}$.
Next, we choose the edge matrix hyperprior.
This matrix can be seen as the adjacency matrix of the multigraph connecting the blocks together.
In Ref.~\cite{peixoto2017nonparametric}, a hierarchical SBM was used as a prior for the edge matrix where each level came with its own node partition and edge matrix that, in turn, can also be modeled by a SBM, and so on until only one block remains.
Here, we focus on the simpler version of this scheme, where the edge matrix prior is simply given by a multigraph ER model with $b$ nodes:
\begin{equation}
    P(\vec{e}|E=e, \vec{b}=\vec{\beta}) = \mutilsetcoeff{\frac{b(b+1)}{2}}{e}^{-1}\,,
\end{equation}
where, again, $b$ is the number of blocks in $\vec{\beta}$.
For the node partition hyperprior, we choose a non-informative uniform distribution on all partitions with $B$ non-empty blocks:
\begin{equation} 
    P(\vec{b}|B=b) = \binom{N-1}{b-1}^{-1}\,,
\end{equation}
which counts the number of possible arrangements of $N$ nodes into $b$ non-empty groups.
Likewise, we choose a non-informative uniform hyperprior over the number of non-empty blocks $B$:
\begin{equation}
    P(B) = N^{-1}\,.
\end{equation}

\section{Markov chain likelihoods}
\label{app:markov-chain}

Throughout the paper, we consider likelihoods where the observations are time series of binary variables for each node, denoted $X=(\vec{X}_1, \vec{X}_2, \dots, \vec{X}_T)$, $\vec{X}_t\in\{0, 1\}^N$ for every $t$.
These data models are based on Markov chains, where the state $X_{t+1}$ at time $t+1$ is conditioned on every previous state except the previous one $\vec{X}_{t}$ at time $t$, that is
\begin{equation}
  P(X|G) = P(\vec{X}_1)\prod_{t=1}^{T} P(\vec{X}_{t+1}|\vec{X}_t, G)\,,
\end{equation}
where $P(\vec{X}_1)$ is the probability distribution of the initial state, and $P(\vec{X}_{t+1}|\vec{X}_t, G)$ is the transition probability from $\vec{X}_t$ to $\vec{X}_{t+1}$.
The type of Markov chains we are interested in are graphical models~\cite{edwards2012introduction}, meaning that the transition probability for a single node $i$ only depends on the previous state of its neighbors including itself:
\begin{equation}
  P(\vec{X}_{t+1}|\vec{X}_t, G=g) = \prod_{i=1}^N P(X_{i, t+1}|X_{i,t}, X_{\mathcal{N}_i, t})\,,
\end{equation}
where $X_{\mathcal{N}_i,t} \equiv (X_{j, t})_{j\in\mathcal{N}_i}$ contains the state of the neighbors $\mathcal{N}_i$ of node $i$ in the graph $g$.
For time homogeneous Markov chains, the transition probability of a node $i$ is expressed in terms of its number of active neighbors $n_{i,t}$, inactive neighbors $m_{i,t}$ and some set of parameters $\varphi$.
We denote the activation and deactivation probabilities $\alpha(n_{i,t}, m_{i,t}, \varphi)$ and $\beta(n_{i,t}, m_{i,t}, \varphi)$, respectively.
Putting everything together, the transition probability of the Markov chain is given by
\begin{widetext}
  \begin{equation}\label{eq:markov-chains}
      \begin{split}
          P\left(\vec{X}_{t + 1}=\vec{y}|\vec{X}_{t}=\vec{x}, G, \phi=\varphi \right) = \prod_{i=1}^N \bigg\{
          &\big[\alpha(n_{i,t}, m_{i,t}, \varphi)\big]^{(1 - x_{i}) y_{i}}
          \big[1 - \alpha(n_{i,t}, m_{i,t}, \varphi)\big]^{(1 - x_{i}) (1 - y_{i})} \\
          &\big[\beta(n_{i,t}, m_{i,t}, \varphi)\big]^{x_{i} (1 - y_{i})}
          \big[1 -\beta(n_{i,t}, m_{i,t}, \varphi)\big]^{x_{i} y_{i}}
          \bigg\}\,.
      \end{split}
  \end{equation}
\end{widetext}

We allow the inactive nodes to spontaneously activate with probability $\alpha_0$, and spontaneously deactivate with probability $\beta_0$. 
Denoting $\tilde{\alpha}$ and $\tilde{\beta}$ the activation and deactivation probabilities without spontaneous activation, respectively, we obtain
\begin{align}
    \alpha(n,m,\varphi) &= (1 - \alpha_0) \tilde{\alpha}(n,m,\varphi) + \alpha_0\,, \\
    \beta(n,m,\varphi) &= (1-\beta_0) \tilde{\beta}(n,m,\varphi) + \beta_0\,.
\end{align}
In general, we fix $\alpha_0 = \beta_0 = 0$ for the synthetic experiments, and infer them in Sec.~\ref{sec:reconstructability-brain-networks}.

\begin{table}[t]
  \centering
  \setlength\tabcolsep{6pt}
  \begin{tabular}{c | c c c}
      \hline \hline
      Dynamics & $\phi$ & $\tilde{\alpha}(n, m)$ & $\tilde{\beta}(n, m)$\\ \hline \\
  Glauber~\cite{glauber1963time} & $J$ & $\sigma\qty(2J(n - m))$ & $\sigma\qty(2J(m - n))$ \\ \\
      SIS~\cite{pastor2015epidemic} & $(\beta, \lambda)$ & $1 - \left(1 - \frac{\lambda}{\beta}\right)^m$ & $\beta$\\ \\
      Voter~\cite{clifford1973model} & $\varnothing$ & $\frac{m}{n + m}$ & $\frac{n}{n + m}$ \\ \\
      Cowan~\cite{cowan1990stochastic}& $(a, \beta, \mu, \nu)$ & $\sigma\qty(a( \nu m - \mu))$ & $\beta$ \\ \\ \hline\hline
  \end{tabular}
  \caption{Activation and deactivation probability functions for the likelihoods used in this paper, where $n$ corresponds to the number of inactive neighbors whose states are $0$, and $m$ corresponds to the number of active neighbors whose states are $1$. We define $\sigma(x) = [\exp(-x) + 1]^{-1}$ as the sigmoid function. 
  }
  \label{tab:dynamics}
\end{table}

Table~\ref{tab:dynamics} presents the activation and deactivation probability functions for four different processes used in various contexts.
The Glauber dynamics is a spin model that describe the time-reversible evolution of magnetic spins ($0$ or $1$) aligning in a crystal.
In this model, the nodes are connected through their neighbors via a coupling constant $J$, that modulates the probability of a node to align with its neighbors.
The susceptible-infected-susceptible (SIS) dynamics is a canonical model of epidemic spreading, where the nodes are either susceptible ($0$) or infected ($1$), and has often been used to model disease with short immunity after recovery, similar to influenza-like disease~\cite{anderson1992infectious}.
Susceptible (or inactive) nodes get infected by each of their infected (active) first neighbors, with a constant transmission probability, and recover from the disease with a constant recovery probability.
The Voter dynamics model the adoption of opinions; A node randomly selects the opinion (two opinions, $0$ or $1$, are considered) of one of its neighbors.
The Cowan dynamics is a model of neural activity of biological neural networks, where the nodes---referred to as neurons---are either active ($1$) or inactive ($0$), and has been used to model the dynamics of single neurons or neuronal populations~\cite{cowan1990stochastic, Painchaud2022}.
Inactive neurons fire---i.e., become active---if their input current, coming from their firing neighbors, is above a given threshold.

The parameters $\phi$ of these models are fixed except in Sec.~\ref{sec:reconstructability-brain-networks} where they are inferred by sampling from the joint posterior $P(G, \phi, \theta| X)$.
In these experiments, we use non-informative uniform prior densities for all parameters in $\phi$, and we constrain their value in finite intervals.
For probability parameters, such as $\alpha_0$ and $\beta$ for the SIS and Cowan models, the prior density is $\rho(\phi) = 1$.
For positive unbounded parameters, such as $J$ for the Glauber model, and $\mu$ and $\nu$ for the Cowan model, we set the maximum value to $10$ such that $\rho(\phi) = \frac{1}{10}$.
Note that we fix $a = 1$ in the case of the Cowan model and $\beta_0 = 0$ for the SIS and the Cowan models in Sec.~\ref{sec:reconstructability-brain-networks}, without loss of generality since they are redundant parameters that may lead to non-identifiability issues.

\section{Heuristic reconstruction algorithms}
\label{app:heuristics}

We consider three heuristic reconstruction approaches in this paper: the correlation matrix method~\cite{kramer2009network}, the Granger causality method~\cite{schreiber2000measuring}, and the transfer entropy method~\cite{seth2005causal}.
The technical details can be found in Ref.~\cite{desrosiers2016network}, and we used the implementations of the \texttt{netrd} package~\cite{mccabe2020netrd}.

These techniques compute a score matrix $S$, such that $S_{ij}$ for each pair of nodes $(i,j)$ correlates with probability that an edge exists between them.
For the correlation matrix method, this score is the autocorrelation coefficient of the Markov chain:
\begin{equation}\label{eq:cross-corr}
    S_{ij} = \frac{C_{ij}}{\sigma_i \sigma_j}\,,\quad C_{ij} = \frac{1}{T-1}\sum_{t=1}^T(X_{i,t} - \bar{X}_i) (X_{j,t} - \bar{X}_j)\,,
\end{equation}
where $\bar{X}_i = \frac{1}{T}\sum_{t=1}^T X_{i,t}$ and $\sigma_i^2 = \frac{1}{T-1}\sum_{t=1}^T (X_{i,t} - \bar{X}_i)^2$. 
The Granger causality method tests the hypothesis that the prediction of the time series of a single node $i$ using a linear auto-regressive model is improved by including the time series of node $j$.
Specifically, it evaluates the statistical significance of error variances to determine if including node $j$'s time series reduces prediction error of $i$'s time series.
This statistical tests is performed using the $F$-statistic:
\begin{equation}
    S_{ij} = \frac{\Sigma_{ij}}{\Sigma_i}\,,
\end{equation}
where $\Sigma_i$ is the error variance of the auto-regressive model of $i$, and $\Sigma_{ij}$ is the error variance of the other model that also includes $j$.
In the transfer entropy method, the score is given by the transfer entropy from the time series of $j$ to the time series of $i$:
\begin{align}
    S_{ij} = T_{X_j\to X_i}\,,
\end{align}
where
\begin{equation}
    T_{X_j \to X_i} = H(X_{i,t+1}|X_{i,t}) - H(X_{i,t+1}|X_{i,t}, X_{j,t})\,.
\end{equation}
The entropies involved in the computation of $T_{X_j\to X_i}$ are evaluated by estimated the probabilities $P(X_{i,t}|X_{i,t-1})$ and $P(X_{i,t}|X_{i,t-1}, X_{j,t-1})$ with the corresponding frequency observed in the time series itself.

\section{Relationship between the posterior loss and the reconstructability}
\label{app:posterior-loss}

In this section we show that the posterior loss is related to the reconstructability, under certain conditions.
Let $(g^*, x^*)$ be generated by the TDG model $M^*=(G^*, X^*)$, and $M=(G,X)$ be the reconstruction model.
We define $p_i(x) = P(G=g_i|X=x)$ be the posterior probability of the graph $g_i$ given some observation $x$.
We also denote $\vec{p}(x) = \big(p_1(x), p_2(x), ..., p_{|\mathcal{G}|}(x)\big)$ the vector of the posterior probabilities of all graphs in $\mathcal{G}$.
The posterior loss $L\big(\vec{y}, \vec{p}(x)\big)$ measures the accuracy of the posterior probabilities $\vec{p}(x)$ at predicting the correct labels $\vec{y}$, where $y_i = \delta(g^*, g_i)$ is a one-hot encoding of the true graph $g^*$ using a Kronecker delta.
It is defined as
\begin{equation}\label{eq:ce-loss1}
  L\big(\vec{y}, \vec{p}\big) = -\sum_{i=1}^{|\mathcal G|} y_i \log p_i(x)\,.
\end{equation}
We also write $L\big(\vec{y}, \vec{p}^*\big)$ the posterior loss of the TDG model $M^*$, such that $\vec {p}^*$ is its corresponding posterior probability vector.
Rewriting the posterior loss in terms of the posterior probability, we simply get
\begin{equation}
  L(\vec{y}, \vec{p}) = -\log P(G=g^*|X=x^*)\,.
\end{equation}
When the posterior probability factors with respect to the edges and the graphs do not contain multiedges, i.e.,
\begin{equation}\label{eq:posterior-factorization}
    P(G=\vec{a}|X=x) = \prod_{i<j} \pi_{ij}(x)^{a_{ij}} \Big(1 - \pi_{ij}(x)\Big)^{1 - a_{ij}}\,,
\end{equation}
the posterior loss is given by Eq.~\eqref{eq:ce-loss}.

The posterior loss averaged over the graph and data generated by $M^*$ is
\begin{align}
  \expect[X^*,G^*]{L(\vec{y}, \vec{p})} &\approx -\expect[X^*,G^*]{\log P(G=G^*|X=X^*)}\,,
\end{align}
where equality is achieved when the posterior distribution $P(G^*|X^*)$ truly factors as in Eq.~\eqref{eq:posterior-factorization}.
Hence, when $M$ and $M^*$ are equal in distribution, $\expect[X^*,G^*]{L(\vec{y}, \vec{p}^*)} \approx H(G^*|X^*)$.
Furthermore, the expected posterior loss is linearly related to the reconstructability, with a proportionality factor given by the entropy of $G^*$:
\begin{equation}
  \expect[X^*,G^*]{L(\vec{y}, \vec{p}^*)} \approx H(G^*)\Big[1 - \Psi^*\Big]\,.
\end{equation}

\section{Bounds of the information gain}
\label{app:information-gain}
In this section, we show that the information gain is non-negative and bounded by the CE between the posterior and the prior.
Recall that the information gain is given by Eq.~\eqref{eq:infogain-1} (equivalently Eq.~\eqref{eq:infogain-2}):
\begin{equation*}
  \mathcal{I}_M(x) = \expect[G|X=x]{\log \frac{P(G|X)}{P(G)}}\,.
\end{equation*}
Jensen's inequality states that for any random variable $Y$ and any convex function $\xi$,
\begin{equation}
  \xi\left(\expect{Y}\right) \leq \expect{\xi(Y)}\,.
\end{equation}
Given that $\xi = - \log$ is a convex function, the information gain can be bounded using Jensen's inequality:
\begin{equation}
  \mathcal{I}_M(x) \geq -\log\expect[G|X=x]{ \frac{P(G)}{P(G|X=x)}}\,.
\end{equation}
Simplifying the right-hand side yields
\begin{align*}
  \mathcal{I}_M(x) \geq&  -\log\left(\sum_{g\in\mathcal{G}}P(G|X=x)\frac{P(G)}{P(G|X=x)} \right)\\
  &= -\log(1) = 0,
\end{align*}
the information gain lower bound $\mathcal{I}_M(x) \geq 0$ for all $M$ and $x\in\mathcal{X}$.

The information gain can also be written as
\begin{equation}\label{eq:infogain-3}
  \mathcal{I}_M(x) = \mathcal{H}\Big(P(G|X=x), P(G)\Big) - H(G|X=x)\,,
\end{equation}
where $\mathcal{H}(p, q) = -\sum_x p(x) \log q(x)$ is the CE between two distributions $p$ and $q$, and 
\begin{equation}
    H(G|X=x) = -\expect[G|X=x]{\log P(G|X=x)}\,
\end{equation}
is the point-wise entropy of the graph posterior distribution for the observation $x$.
The information gain is maximized when $H(G|X=x)$ is minimized, i.e., zero.
The remaining term---the cross-entropy---is thus the upper bound of the information gain:
\begin{equation}
  \mathcal{I}_M(x) \leq -\expect[G|X=x]{\log P(G)} = \Lambda_M(x)\,.
\end{equation}

\section{Numerical approximations of the mutual information}
\label{app:numerical-techniques}
The mutual information $\mi$ and information gain $\mathcal{I}_M(x)$ are generally intractable. 
Their intractability stems from the evaluation of the posterior, which requires computing of the evidence, denoted by $\zeta_M(x) = P(X=x)$:
\begin{equation}\label{eq:evidence}
    \zeta_M(x) = \sum_{g\in\mathcal{G}} P(G=g) P(X=x| G=g)\,.
\end{equation}
Indeed, there are potentially an exponential number of terms in this sum that need to be evaluated.
Moreover, if $M$ involves hyperparameters $\theta$ or parameters $\phi$, they must also be marginalized to find the evidence.
Fortunately, the evidence probability can be estimated efficiently using Monte Carlo techniques as described in this section.
Note that we focus on the mutual information computation, but the same techniques can be applied to the information gain.

\subsection{Graph enumeration approach}
For sufficiently small random graphs ($N\approx 5$), the evidence probability can be computed by enumerating all graphs of $\mathcal{G}$ and by adding explicitly each term of Eq.~\eqref{eq:evidence}. 
Using the law of large numbers, we can estimate the mutual information
\begin{equation}
    \begin{split}
        \mi \simeq \frac{1}{K} \sum_{k=1}^K &\bigg[\log P\Big(X=x^{(k)}|G=g^{(k)}\Big) \\&
        - \log P\Big(X=x^{(k)}\Big)\bigg]\,,
    \end{split}
\end{equation}
where $(x^{(k)}, g^{(k)})_{k=1..K}$ are pairs of time series and graph sampled from $(X, G)$ for $M$, the Bayesian generative model.
The variance of this estimator scales with $K^{-1/2}$.

\subsection{Variational mean-field approximation}
This approach is based on Ref.~\cite{murphy2024duality} which uses a variational mean-field approximation to estimate the posterior probability instead of the evidence probability. 
The variational mean-field (MF) approximation assumes the conditional independence of the edges. 
For simple graphs, the MF posterior is
\begin{equation}\label{sieq:mf_posterior}
    P_\mathrm{MF}(G=\vec{a}|X=x) = \prod_{i\leq j} [\pi_{ij}(x)]^{a_{ij}}\, [1 - \pi_{ij}(x)]^{1 - a_{ij}}\,,
\end{equation}
where $\pi_{ij}(x)\equiv P(A_{ij}=1|X=x)$ is the marginal conditional probability of existence of the edge $(i, j)$ given $x$. 
For multigraphs, we obtain a similar expression involving a probability $\pi_{ij}(m\mid x) = P(A_{ij}=m|X=x)$ that there are $m$ multiedges between $i$ and $j$. 
In this case, the MF posterior becomes
\begin{equation}
    P_\mathrm{MF}(G=\vec{a}|X=x) = \prod_{i<j} \pi_{ij}(a_{ij}\mid x)\,.
\end{equation} 
By the conditional independent between the edges~\cite[Theorem 2.6.5]{cover2006elements}, the MF approximation is a lower bound of the posterior entropy
\begin{equation}\label{sieq:mf_upperbound}
    H(G|X) \leq -\expect[X,G]{\log P_\mathrm{MF}(G|X)}\,.
\end{equation}
As for the graph enumeration approach, we compute the MF estimator of the mutual information with the Monte Carlo estimator
\begin{equation}\label{sieq:mf_estimator}
    \begin{split}
        I(G ; X) \gtrsim \frac{1}{K}\sum_{k=1}^K \Big[&\log P_\mathrm{MF}\Big(G=g^{(k)}|X=x^{(k)}\Big) \\&- \log P\Big(G=g^{(k)}\Big)\Big]\,.
    \end{split}
\end{equation}
The posterior probability $P_\mathrm{MF}\Big(G=g^{(k)}|X=x^{(k)}\Big)$ is also found using the law of large numbers: $\pi_{ij}(x)$ is estimated as the proportion of graphs that contain the edge $(i,j)$ in a sample of the posterior.
An analogous estimation is made in the multigraph case, where $\pi_{ij}(a_{ij}\mid X)$ is the proportion of graphs that contain $a_{ij}$ edges between $i$ and $j$ in the sample. 
Although Eq.~\eqref{sieq:mf_estimator} is a biased estimator of the mutual information, it was shown in Ref.~\cite{murphy2024duality} that the bias is generally small, especially for large networks. 

\subsection{Graph evidence estimation for the stochastic block model}
\label{sub:sbm-graph-evidence}
Using the stochastic block model (SBM) as the prior for our reconstruction model and for estimating the mutual information is challenging. 
Indeed, computing the graph entropy $H(G)$ requires that we marginalize the partition out of the prior probability
\begin{equation}
    P(G) = \sum_{\vec{b}} P(G, \vec{b})\,,
\end{equation}
which is intractable, but can be estimated.
In Ref.~\cite{peixoto2021revealing}, the author proposes a way to estimate the probability $P(\vec{b}|G)$ of partition given $G$---i.e., the posterior of a Bayesian model for community detection---by sampling a set of $M$ partitions from it using Markov chain Monte Carlo (MCMC).
The complete procedure is complex and involves aligning the sampled partitions, identifying aligned partition clusters and estimating the node marginal partition distribution $P(b_i=r|G) \equiv \pi_{i,r}(G)$ that node $i$ is in group $r$---we refer to the original paper for technical details.
To evaluate the graph marginal log probability, we first notice that 
\begin{align}
    \log P(G) &= \expect[\vec{b}|G]{\log P(G)}\,, \notag\\
    &= \expect[\vec{b}|G]{\log P(G, \vec{b})} - H(\vec{b}|G)\,.
    \label{eq:evidence_decomposition_entropy_sbm}
\end{align}
Given that we know the joint probability $P(G,\vec{b})$, the goal is then to estimate the partition entropy $H(\vec{b}|G)$.
In Ref.~\cite{peixoto2021revealing}, they propose a standard mean-field estimator:
\begin{equation}\label{eq:partition-posterior}
    P_\mathrm{MF}(\vec{b}|G) = \prod_{i=1}^N\pi_{i,b_i}(G)\,,
\end{equation}
where the marginal probabilities $\pi_{i,r}(G)$ can be estimated by the fraction of sampled relabeled partitions where node $i$ is in block $r$.
The mean-field estimator of the partition entropy is then
\begin{equation}
    H_\mathrm{MF}(\vec{b}|G) = -\sum_{i=1}^N \sum_{r=1}^{B_{\max}} \pi_{i, r}(G) \log \pi_{i,r}(G) \,,
\end{equation}
where $B_{\max}$ is typically chosen to be equal to $N$, as there can be at most $N$ non-empty groups.
Also, note that $H_\mathrm{MF}(\vec{b}|G)\geq H(\vec{b}|G)$, since by factoring as in Eq.~\eqref{eq:partition-posterior} we assume that the node memberships are conditionally independent, which has the effect of increasing the entropy~\cite{cover2006elements}.
Finally, the graph evidence entropy can be estimated using that mean-field estimator as follows:
\begin{equation}
    H(G) \geq H(G, \vec{b}) - H_\mathrm{MF}(\vec{b}|G)\,,
\end{equation}
which in turn constitutes a lower bound of $H(G)$.

\subsection{Evidence estimation for model selection}
The estimation of the evidence log probability relies on the previously discussed techniques for evaluating the posterior probability.
Using the same approach as for Eq.~\eqref{eq:evidence_decomposition_entropy_sbm}, we obtain
\begin{equation}
    \log \zeta(x)= \expect[G|X=x]{\log P(X, G)}- H(G|X=x)\,,
\end{equation}
which follows from the fact that $\log P(X) = \log P(X, G) - \log P(G|X)$.
Hence, we build an estimator by sampling from the posterior $K$ graphs $g^{(k)}$ given $x$
\begin{equation}
\begin{split}
    \log \zeta(x) \simeq &\frac{1}{K}\sum_{k=1}^K \bigg[ \log P(G=g^{(k)},X=x) \\&- H(G|X=x) \bigg] \,.
\end{split}
\end{equation}
By replacing $H(G|X=x)$ with a MF estimator of the posterior entropy, e.g. using Eq.~\eqref{sieq:mf_posterior} for simple graphs, we asymptotically get a lower bound of the evidence log probability:
\begin{equation*}
\begin{split}
    \log \zeta(x) \gtrsim &\frac{1}{K}\sum_{k=1}^K \bigg[\log P(G=g^{(k)}, X=x)\\ &- \sum_{i<i} h\big(\pi_{ij}(x)\Big)\bigg]\,,
\end{split}
\end{equation*}
where we recall that $h(p) = -p\log p - (1 - p) \log (1 - p)$ is the binary entropy.

When there are parameters $\theta$ and $\phi$ for the graph $G$ and data $X$, respectively, to infer alongside $G$, they must also be marginalized in the calculation of the evidence.
Using a similar strategy as in the case where only $G$ is inferred, we start from
\begin{equation}\label{eq:generalized-log-evidence}
\begin{split}
    \log \zeta(x) = \,\,&\expect[\theta, \phi, G|X=x]{\log P(X, \phi, G, \theta)}\\ &- H(\phi, G, \theta|X=x)\,,
\end{split}
\end{equation}
where we note that the expectation is taken over the complete joint posterior distribution $P(\phi, G, \theta|X=x)$.
While the estimation of the first term is performed as previously, that of the second term, i.e., the posterior joint entropy, is more tricky.
To build an estimator, we take advantage of the fact that the variables $\phi$, $G$ and $\theta$ are conditionally dependent in a specific way $\theta\rightarrow G \rightarrow X \leftarrow \phi$, as we previously have pointed out.
This means that the posterior joint entropy can be factored in the following way
\begin{align}
    H(\phi, G, \theta|X) &= H(\phi|G, \theta, X) + H(\theta|G,X) + H(G|X)\notag\\
    &= H(\phi|X) + H(\theta|G) + H(G|X)\,,
\end{align}
where $H(\phi|X) = H(\phi|G, \theta, X)$ and $H(\theta|G,X) = H(\theta|G)$, by virtue of the facts that $\phi$ is conditionally independent of $G$ and $\theta$, and that $\theta$ is conditionally independent of $X$.
For evaluating $H(\phi|X)$, since $\phi$ are continuous random variables, we estimate the posterior density with kernel density estimation (KDE) with a Gaussian kernel and estimate the differential entropy from the estimated density.
In the case of $H(\theta|G)$, this term only concerns the SBM prior in our experiments, where $\theta$ are discrete variables $\vec{b}$ where $b_i$ denotes the membership of node $i$ to a group.
We use the procedure described in Sec.~\ref{sub:sbm-graph-evidence} to estimate $H(\vec{b}|G)$.

\section{Markov chain Monte-Carlo algorithm}
\label{app:mcmc-algo}

To sample from the posterior distribution, we use a Markov chain Monte Carlo (MCMC) algorithm. 
Starting from a graph $g$, we propose a move to graph $g'$, according to a proposition probability $P(G'=g'| G=g)$, and accept it with the Metropolis-Hastings probability:
\begin{equation}\label{eq:accept-prob}
    \min\left(1,  e^{-\log\Delta}\frac{P(G'=g|G=g')}{P(G'=g'|G=g)}\right)\,,
\end{equation}
where $\Delta = \frac{P(G=g')P(X=x|G=g')}{P(G=g)P(X=x|G=g)}$ is the ratio between the posterior probabilities of $g$ and $g'$. 
This ratio can be computed efficiently in $\bigo{T}$ by keeping in memory, for each node $i$ and time $t$, the number of inactive neighbors $n_{i,t}$ and the number of active neighbors $m_{i,t}$ (see Refs.~\cite{peixoto2019network,murphy2024duality}).
Equation~\eqref{eq:accept-prob} allows to sample from the posterior distribution $P(G\mid X)$ without the requirement to compute the intractable normalization constant $P(X)$.

We use two types of graph move propositions: double-edge swaps and hinge flips~\cite{coolen2017mcmc}.
Double-edge swaps consists in selecting two edges at random, breaking them into two pairs of stubs and reconnecting the stubs to create two new edges.
This type of move leaves the degree sequence and total edge count unchanged.
Hinge flips consist in selecting an edge and a node at random, and reconnecting the edge to this node by detaching it from one of its end.
Unlike double-edge swaps, hinge flips do not preserve the degree sequence.
There are many considerations to take when implementing these moves and computing their proposal probabilities; we refer to Refs.~\cite{coolen2017mcmc,fosdick2018configuring,murphy2024duality} for technical details.

For most of our numerical experiments, the total number of edges is fixed.
At each proposition, we randomly select to perform a double-edge swap or a hinge flip with equal probability.
We found that by doing this, the mixing time was significantly improved.

This sampling scheme can be generalized when additional hyperparameters $\theta$ of the graph prior or parameters $\phi$ from the likelihood must be inferred as in Sec.~\ref{sec:reconstructability-real-systems}.
We consider a Gibbs sampling scheme where each random variable $G$, $\theta$ and $\phi$ is sampled sequentially and conditioned on the others.
In all cases, the acceptance probability follows Eq.~\eqref{eq:accept-prob}, where $G$ is replaced by either $\theta$ or $\phi$ when these parameters are sampled.
In this paper, only the SBM among the considered graph models contains parameters of the type of $\theta$.
To sample from these, we use the same procedure as in Ref.~\cite[Sec.~VI]{peixoto2017nonparametric}---we refer to it for further detail.
The data models considered contains many parameters that we would like to infer, for example, the infection and recovery probabilities, $\lambda$ and $\beta$, respectively, in the SIS.
These parameters are real number constrained within an interval (for instance, $[0, 1]$ for the recovery probability $\beta$); Hence, any proposed move where $\varphi$ falls outside of this interval is rejected.
We propose moves drawn from a Gaussian distribution with density
\begin{equation}
    p(\varphi'|\varphi) \propto \exp\left[-\frac{(\varphi' - \varphi)^2}{2\sigma^2}\right]\,.
\end{equation}
In Sec.~\ref{sec:reconstructability-real-systems}, we fix $\sigma=0.1$.

\section{Reconstructability of graph models with delta distribution}
\label{app:reconstructability-delta}
Suppose $X$ is generated using a single graph $g^*$. 
If we were to observe many realizations of $X$ with the single graph $g^*$, the graph prior of the TDG would be $P(G=g) = \delta(g, g^*)$ and the evidence of this process would be exactly equal to the likelihood of the TDG process, denoted $p^*(X) \equiv P(X|G=g^*)$. 
As a result, the mutual information and the entropy of $G$ would both be zero, and so the reconstructability would be undefined.

To bypass this problem, suppose that the graph generating model is instead parametrized by a probability $\epsilon$ such that $G$ yields $g^*$ with probability $1-\epsilon$ and the others uniformly, that is,
\begin{equation}
    P(G=g) = \begin{cases} (1 - \epsilon) & \text{if } g=g^*, \\ \frac{\epsilon}{Z} &\text{otherwise},\end{cases}
\end{equation}
where $Z = |\mathcal{Z}|$ such that $\mathcal{Z} = \set{g\in\mathcal{G} : g\neq g^*}$ is the set of graphs different from $g^*$.
Then, by taking the limit when $\epsilon \to0$, we recover the scenario where the graph generating model is a Kronecker delta distribution.

Let us investigate the scaling of $H(G)$ and $I(X;G)$. First, we have
\begin{align}
    H(G) &= -(1 - \epsilon) \log(1- \epsilon) - \epsilon \sum_{g\in\mathcal{Z}} \frac{1}{Z} \log\frac{\epsilon}{Z} \nonumber\\
    &= h(\epsilon) + \epsilon\log Z \,.\label{eq:entropy_graph_epsilon}
\end{align}
Second, we have the evidence of this joint model:
\begin{align*}
    P(X=x) &= (1 - \epsilon)p^*(x) + \epsilon\sum_{g\in\mathcal{Z}} \frac{P(X=x|G=g)}{Z}\\
    &= (1 - \epsilon) p^*(x) + \epsilon q(x)\,,
\end{align*}
where $q(x) = \sum_{g\in\mathcal{Z}} \frac{P(X=x|G=g)}{Z}$ is the evidence of $x$ in the complementary model for which the only possible graphs are those in $\mathcal{Z}$. Then, the reconstruction entropy $H(G|X)$ is evaluated as follows:
\begin{widetext}
\begin{align*}
    H(G|X) = & -\sum_x \bigg[ P(X=x,G=g^*)\log P(G=g^*|X=x) + \sum_{g\in \mathcal{Z}} P(X=x,G=g)\log P(G=g|X=x) \bigg] \\
    = & -\sum_x \bigg[ P(X=x|G=g^*)P(G=g^*)\log \frac{P(X=x| G=g^*)P(G=g^*)}{P(X=x)} \\
    & \qquad \qquad + \sum_{g\in \mathcal{Z}} P(X=x|G=g)P(G=g)\log  \frac{P(X=x| G=g)P(G=g)}{P(X=x)} \bigg]\\
    =&-\sum_x \bigg[ (1 - \epsilon)p^*(x)\log \left[\frac{(1 - \epsilon)p^*(x)}{(1 - \epsilon)p^*(x) + \epsilon q(x)}\right] \\
    &\qquad \qquad +\epsilon \sum_{g\in\mathcal{Z}} Z^{-1}P(X=x|G=g)\log \left[\frac{\epsilon Z^{-1} P(X=x|G=g)}{(1 - \epsilon)p^*(x) + \epsilon q(x)}\right] \bigg]\\
    =&-\sum_x \bigg[ (1 - \epsilon)p^*(x)\log (1-\epsilon) + (1 - \epsilon) p^*(x)\log \left[\frac{p^*(x)}{(1 - \epsilon)p^*(x) + \epsilon q(x)}\right]\\
    &\qquad \qquad +\epsilon \sum_{g\in\mathcal{Z}} Z^{-1}P(X=x|G=g)\log \frac{\epsilon}{Z} + \epsilon \sum_{g\in\mathcal{Z}} Z^{-1}P(X=x|G=g)\log \left[\frac{ P(X=x|G=g)}{(1 - \epsilon)p^*(x) + \epsilon q(x)}\right] \bigg]\\
    =&\, h(\epsilon) +\epsilon \log Z - (1 - \epsilon) A  - \epsilon B + \epsilon H(X|\bar{G}),
\end{align*}
\end{widetext}
where
\begin{align*}
    A &= -\sum_x p^*(x) \log\left[1 + \epsilon \left(\frac{q(x)}{p^*(x)} - 1\right)\right]\,, \\
    B &= -\sum_x q(x) \log\left[(1 - \epsilon)p^*(x) + \epsilon q(x)\right]\,,
\end{align*}
and $\bar{G}$ denotes the random graph uniformly distributed over the complementary set $\mathcal{Z}$.
Recalling Eqs.~\eqref{eq:mutual-information} and \eqref{eq:entropy_graph_epsilon}, we conclude that 
\begin{align*}
    I(X;G) = (1 - \epsilon) A  + \epsilon B - \epsilon H(X|\bar{G}).
\end{align*}

By developing the logarithms as $\log (1 + x) = x + \mathcal{O}(x^2)$, we can show easily that the leading term of $A$ is of second order in $\epsilon$ (no constant or linear terms) and the leading terms of $B$ is
\begin{align*}
    B = -\sum_x q(x) \left[\log p^*(x) + \epsilon\left(\frac{q(x)}{p^*(x)} - 1\right) + \mathcal{O}(\epsilon^2)\right]\,.
\end{align*}
This leaves us with 
\begin{align*}
I(X;G) = \epsilon\Big(-H(X|\bar{G}) - \sum_{x} q(x) \log p^*(x)\Big) + \mathcal{O}(\epsilon^2).
\end{align*}
Also, from the above equation for $H(G)$, we have that the leading terms are $H(G) = \epsilon (\log\epsilon^{-1} + 1 + \log Z) + \mathcal{O}(\epsilon^2)$. Consequently, the reconstructability, being the ratio of $I(X;G)$ and $H(G)$, approaches zero as $\epsilon\to 0$ with leading term $\bigo{\frac{1}{\log \epsilon^{-1}}}$.

\section{Inference of brain networks}
\label{app:inference-brain}

\begin{figure}
    \centering
    \includegraphics[width=0.45\textwidth]{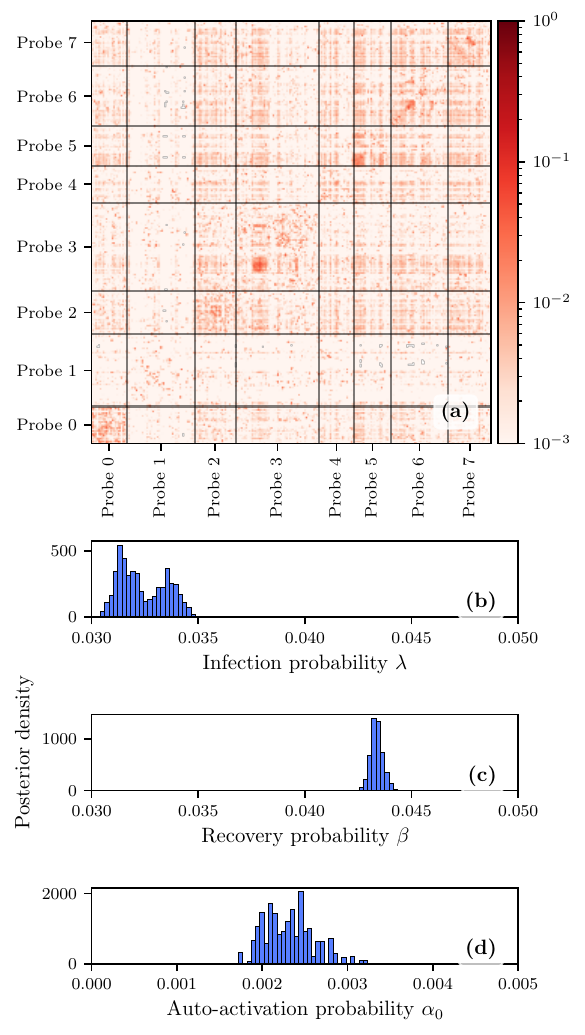}
    \caption{Posterior of the maximum evidence model (SIS model with SBM prior): (a)  posterior probability matrix of the edge occupancy, (b) histogram of the infection probability, (c) the recovery probability and (d) the auto-activation probability. In (a), each entry of the matrix represents the number of times the edge has been sampled, among the 8000 posterior samples. Also, we highlight the probe partition of the graph using deemed black separation lines.}
    \label{fig:posterior-characterize}
\end{figure}

\begin{figure}
    \centering
    \includegraphics[width=0.45\textwidth]{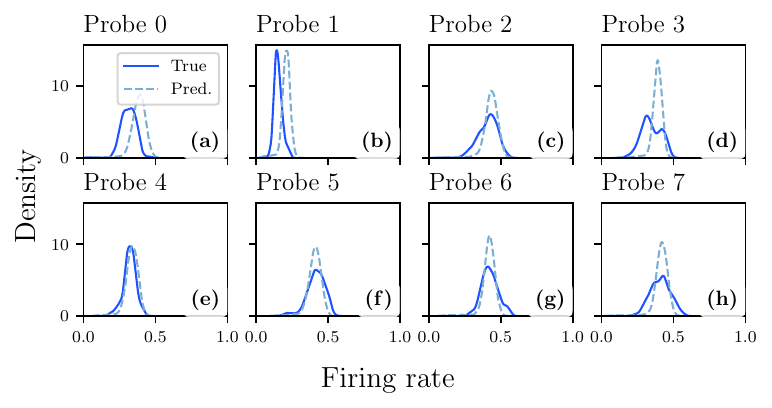}
    \includegraphics[width=0.45\textwidth]{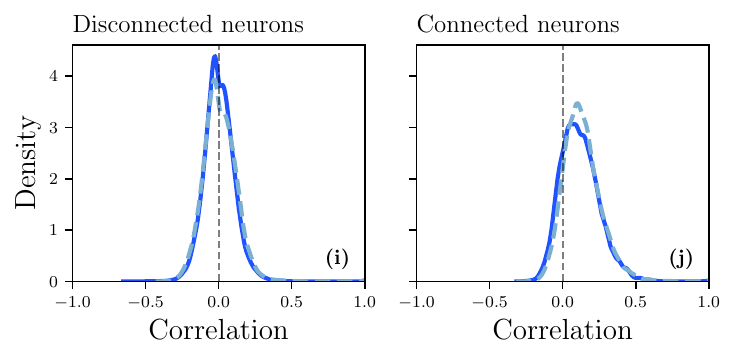}
    \caption{Posterior predictive checks of the maximum evidence model (SIS model with SBM prior), showing Gaussian kernel density estimations of the distributions of (a--h) firing rates (i--j) correlation. 
    Panels (a--h) show the firing rate probability density for each probe.
    In panels (i--j), we show the probability density of the correlation coefficients (Eq.~\eqref{eq:cross-corr}) between neurons that are connected [panel (i)] and disconnected [panel (j)] in the posterior graph.
    In all panels, the statistics corresponding to the observed time series [Fig.~\ref{fig:spiking-neurons}(a), labeled "True"] are shown using the solid dark blue lines, while those of the posterior predictions are shown using the dashed light blue line (labeled "Pred."). 
    Also, the predictions are gathered from 100 samples of the model, where each used different parameters and graph jointly sampled from the posterior.} 
    \label{fig:posterior-check}
\end{figure}

In this section, we describe the procedure we used to reconstruct the mouse brain network from Sec.~\ref{sec:reconstructability-brain-networks}.
The raw data is available in~\cite{steinmetz2019eight}, which was originally presented in Ref.~\cite{stringer2019spontaneous}.
We refer to their paper for any technical detail regarding the data collection.

\begin{table}[t]
    \centering
    \begin{tabular}{ll|c c}
    \hline\hline
     &  & Average & Std. Dev.\\ \hline
    Model & Graph prior & & \\ \hline
    \multirow[t]{3}{*}{Cowan} & ER & 1\,296.75 & 77.00 \\
    & UCM & 1\,302.75 & 70.38 \\
     & SBM & 1\,462.00 & 126.03\\ \hline
    \multirow[t]{3}{*}{Glauber} & ER & 18\,098.88 & 53.94\\
    & UCM & 18\,612.38 & 93.50\\
     & SBM & 18\,895.63 & 79.47\\ \hline
    \multirow[t]{3}{*}{\textbf{SIS}} & ER & 1\,388.50 & 28.92\\
    & UCM & 1\,296.63 & 50.15\\
     & \textbf{SBM} & \textbf{1\,722.38} & \textbf{89.98}\\
    \hline \hline
    \end{tabular}
    \caption{Statistics for the number of edges determined from the semi-greedy algorithm for each reconstruction model considered in Sec.~\ref{sec:reconstructability-brain-networks}. The highlighted row (SIS with SBM) corresponds to the maximum evidence model associated with Figs.~\ref{fig:posterior-characterize} and \ref{fig:posterior-check}. The average and standard deviations (std. dev.) are obtained from the 8 parallel chains used for the inference.}
    \label{tab:edge_count_stats}
\end{table}

\subsection{Data preprocessing}
\label{sec:app:data-preprocessing}
This dataset is composed of the spontaneous activity of the brains of three mice (Krebs, Robbins and Waksman) monitored via eight neuropixel probes each.
These probes record the time stamps of each spike of individual neurons in different regions of the brain for a duration of 20 minutes.

For the purpose of the experiment, we choose the Krebs recoding which count 1462 monitored neurons.
First, we discretize time into $10^5$ steps and map each spike time stamp to the correct discrete time interval.
Then, since the time duration of the spikes are not available in the original dataset, we artificially extend the spikes for a random duration, which is exponentially distributed with mean $0.012$ seconds---this value corresponds to an approximate activation duration of 10 time steps.
Finally, we partition the complete time series into 100 segments of equal size (1000 steps).
Figure~\ref{fig:spiking-neurons}(a) corresponds to the first among the 100 segments of the discretized time series.

\subsection{Inference procedure}
The inference procedure is very similar to that presented in Appendix~\ref{app:mcmc-algo}.
We consider a model parametrized by $X$, $G$ and their parameters $\phi$ and $\theta$, respectively.
However, we have an additional limitation: We do not know the number of edges in the graph.
We tried extending our MCMC algorithm by including moves that do not preserve the number of edges---i.e., adding or removing a single edge---, but we found that these attempts suffered from poor mixing time.

To alleviate this problem, we propose to search for the number of edges first, by minimizing the description length $\log P(X, \phi, G, \theta)$~\cite{peixoto2024network}.
We solve this optimization problem using a semi-greedy algorithm where we propose $K$ move candidates, and select the one that locally minimizes the objective.
Like for our MCMC algorithm, we iterate over $G$, $\theta$ and $\phi$ sequentially to locally perform the optimization on each of them independently.
At each step, we sample $10000$ candidates for $G$ and $\theta$, and $10$ for $\phi$.
Once the number of edges has converged, we stop the semi-greedy algorithm and freeze the number of edges.
The MCMC algorithm then proceeds to sample from the posterior with a fixed number of edges.

In Table~\ref{tab:edge_count_stats}, we summarize the results of the semi-greedy search for the number of edges. 
Given that the number of nodes is $1462$, our results show that the inferred networks are surprisingly sparse, except for the Glauber model which inferred one order of magnitude more edges than the Cowan and SIS models.

\subsection{Posterior inspection}
The graph and parameter marginal posteriors of the maximum evidence model are illustrated in Fig.~\ref{fig:posterior-characterize}.
We also include a validation of the posterior on the inference data.
Figure~\ref{fig:posterior-check} shows the posterior predictive check validation, which includes a prediction of the firing rates and the correlation coefficients between connected and disconnected neurons as test quantities.
The inferred graph also allows to reproduce the shape of the cross-correlation density distribution.

\section*{Acknowledgments}
The authors want to thank J.-F. Fortin, N. Doyon and G. Petri for their helpful comments. This work was supported by the Fonds de recherche du Qu\'ebec -- Nature et technologies (AL, VT), the Conseil de recherches en sciences naturelles et en g\'enie du Canada (CM, PD, AA), the Sentinelle Nord initiative funded by the Fonds d’excellence en recherche Apog\'ee Canada (CM, AL, FT, PD, AA), and the Fonds d'accélération des collaboration en santé du Québec -- Alliance Neuro-CERVO (PD, AA). We acknowledge Calcul Qu\'ebec, Digital Research Alliance of Canada and Hectiq.Ai for their technical support and computing infrastructures.

%

\end{document}